\documentclass[usenatbib]{mn2e}

\usepackage{graphicx}
\usepackage{graphics}
\usepackage{color}
\usepackage{url}
\usepackage[caption=false]{subfig}

\newcommand{\name}{ASASSN-14ae}
\newcommand{\galname}{SDSS J110840.11+340552.2}
\newcommand{\swift}{{\it Swift}}
\newcommand{\msun}{\ensuremath{\rm{M}_\odot}}
\newcommand{\lsun}{\ensuremath{\rm{L}_\odot}}

\def \s{\hphantom{1}}
\newcommand{\ion}[2]{{#1~\uppercase \expandafter{\romannumeral #2}}}
\newcommand{\edit}[1]{\textcolor{black}{#1}}

\def\apj{\rm ApJ}
\def\apjl{\rm ApJL}
\def\apjs{\rm ApJS}
\def\aj{\rm AJ}
\def\aa{\rm A\&A}
\def\mnras{\rm MNRAS}
\def\nat{\rm Nature}
\def\pasj{\rm PASJ}
\def\pasp{\rm PASP}
\def\aap{\rm AAP}

\begin{document}

\title[ASASSN-14ae: A TDE at 200~Mpc]{ASASSN-14ae: A Tidal Disruption Event at 200~Mpc}

\author[T.~W.-S.~Holoien et al.]{T.~W.-S.~Holoien$^{1}$, J.~L.~Prieto$^{2,3,4}$, D.~Bersier$^{5}$, C.~S.~Kochanek$^{1,6}$, K.~Z.~Stanek$^{1,6}$,
\newauthor
B.~J.~Shappee$^{1}$, D.~Grupe$^{7,8}$, U.~Basu$^{1,9}$, J.~F.~Beacom$^{1,6,10}$, J.~Brimacombe$^{11}$, 
\newauthor
J.~S.~Brown$^{1}$, A.~B.~Davis$^{10}$, J.~Jencson$^{1}$, G.~Pojmanski$^{12}$ and D.~M.~Szczygie{\l}$^{12}$ \\
  $^{1}$ Department of Astronomy, The Ohio State University, 140 West 18th Avenue, Columbus, OH 43210, USA \\
  $^{2}$ Department of Astrophysical Sciences, Princeton University, 4 Ivy Lane, Peyton Hall, Princeton, NJ 08544, USA \\
  $^{3}$ N\'ucleo de Astronom\'ia de la Facultad de Ingenier\'ia, Universidad Diego Portales, Av. Ej\'ercito 441, Santiago, Chile \\
  $^{4}$ Millennium Institute of Astrophysics \\
  $^{5}$ Astrophysics Research Institute, Liverpool John Moores University, 146 Brownlow Hill, Liverpool L3 5RF, UK \\
  $^{6}$ Center for Cosmology and AstroParticle Physics (CCAPP), The Ohio State University, 191 W. Woodruff Ave., Columbus, OH 43210, USA \\
  $^{7}$ Swift MOC, 2582 Gateway Dr., State College, PA 16802, USA \\
  $^{8}$ Space Science Center, Morehead State University, 235 Martindale Dr., Morehead, KY 40351, USA \\
  $^{9}$ Grove City High School, 4665 Hoover Road, Grove City, OH 43123, USA \\
  $^{10}$ Department of Physics, The Ohio State University, 191 W. Woodruff Ave., Columbus, OH 43210, USA \\
  $^{11}$ Coral Towers Observatory, Cairns, Queensland 4870, Australia \\
  $^{12}$ Warsaw University Astronomical Observatory, Al. Ujazdowskie 4, 00-478 Warsaw, Poland
   }

\maketitle

\begin{abstract}
{\name} is a candidate tidal disruption event (TDE) found at the center of {\galname} ($d\simeq200$~Mpc) by the All-Sky Automated Survey for Supernovae (ASAS-SN). We present ground-based and {\swift} follow-up photometric and spectroscopic observations of the source, finding that the transient had a peak luminosity of $L\simeq8\times10^{43}$~erg~s$^{-1}$ and a total integrated energy of $E\simeq1.7\times10^{50}$ ergs \edit{radiated} over the $\sim5$ months of observations presented. The blackbody temperature of the transient remains roughly constant at $T\sim20,000$~K while the luminosity declines by nearly 1.5 orders of magnitude during this time, a drop that is most consistent with an exponential, $L\propto e^{-t/t_0}$ with $t_0\simeq39$~days.  The source has broad Balmer lines in emission at all epochs as well as a broad \ion{He}{2} feature emerging in later epochs. We compare the color and spectral evolution to both supernovae and normal AGN to show that {\name} does not resemble either type of object and conclude that a TDE is the most likely explanation for our observations. At $z=0.0436$, {\name} is the lowest-redshift TDE candidate discovered at optical/UV wavelengths to date, and we estimate that ASAS-SN may discover $0.1 - 3$ of these events every year in the future.
\end{abstract}


\section{Introduction}
\label{sec:intro}

When a star's orbit brings it within the tidal disruption radius of a supermassive black hole (SMBH), the tidal shear forces become more powerful than the star's self-gravity and the star breaks apart. Roughly half of the mass of the star is ejected while the rest of the stellar material remains bound to the black hole and is accreted. These tidal disruption events (TDEs) result in a short-lived ($t\la1$~yr) accretion flare \citep{lacy82,phinney89,rees88,evans89}. For $M_{BH}\la10^7 {\msun}$, the initial fallback rate is super-Eddington and the eventual rate at which material returns to pericenter becomes a $t^{-5/3}$ power law \citep{evans89,phinney89}. While it is commonly assumed that the resulting luminosity is proportional to this rate of return to pericenter, this is only a crude approximation to the complex physics associated with evolution of the accretion stream \citep{kochanek94}, and the exact return rates depend on, for example, the structure of the star \citep[e.g.,][]{lodato11}.

In the most luminous phases, TDE emission is likely dominated by a photosphere formed in the debris rather than any direct emission from a hot accretion disk \citep{evans89,loeb97,ulmer99,strubbe09}. Only in the late phases, as the debris becomes optically thin, will there be any direct emission from the disk. The exact balance likely depends on the viewing angle, as illustrated by the simulations of \citet{guillochon14}. Observationally, TDEs would be expected to show spectral characteristics and light curve evolution that would distinguish them from both supernovae (SNe) and normal active galactic nuclei (AGN), and the detection and study of TDEs remains a useful avenue for studying the properties of SMBHs despite their low predicted frequency \citep[$(1.5 - 2.0)_{-1.3}^{+2.7} \times 10^{-5}~{\rm yr}^{-1}$ per galaxy;][]{velzen14}, as the light emitted during the TDE flare may be sensitive to the black hole spin and mass \citep[e.g.,][]{ulmer99,graham01}.

At present candidate TDEs can be divided into two observational classes based on the wavelength at which they were discovered. The first consists of those found in UV and X-ray surveys, such as candidates in NGC5905 \citep{komossa99} and IC3599 \citep{grupe95,brandt95}, a candidate in the galaxy cluster A1795 \citep{donato14}, Swift J164449.3+573451 \citep{burrows11,bloom11,levan11,zauderer11}, Swift J0258.4+0516  \citep{cenko12b}, and GALEX candidates D1-9, D3-13, and D23H-1 \citep{gezari08,gezari09}. These typically do not have strong optical emission. The second consists of those found in optical surveys, \edit{including PTF10iya \citep{cenko12a}; SDSS TDE 1 and TDE 2 \citep{velzen11}; PS1-10jh \citep{gezari12b}; PS1-11af \citep{chornock14}; and PTF09ge, PTF09axc, PTF09djl, PTF10iam, PTF10nuj, and PTF11glr \citep{arcavi14}}. While the X-ray candidates currently outnumber the optical and UV candidates, variable AGN activity can also mimic the behavior expected for TDEs in the X-ray, making it difficult to disentangle true TDEs from AGN \citep{velzen11}. Modern wide-field optical transient surveys, such as the All-Sky Automated Survey for Supernovae (ASAS-SN\footnote{\url{http://www.astronomy.ohio-state.edu/~assassin/index.shtml}}; \citealt{shappee13}), the Palomar Transient Factory \citep[PTF;][]{law09}, and the Panoramic Survey Telescope \& Rapid Response System \citep[Pan-STARRS;][]{chambers07} allow for the identification and monitoring of such TDE candidate events on a rapid time cadence across many wavelengths, helping to differentiate true TDEs from AGN and SNe, and \edit{should prove a very useful resource for discovering TDFs in the future.}

\begin{figure}
\centering
\includegraphics[width=\linewidth]{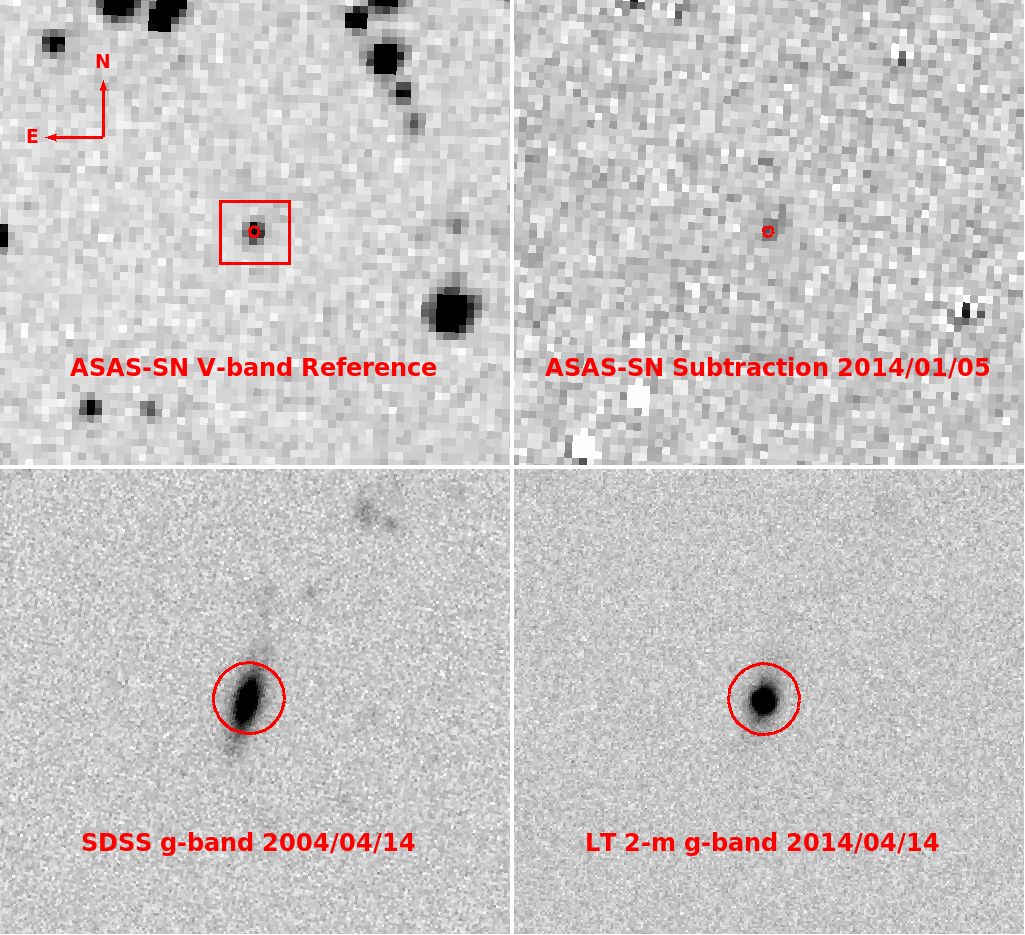} 
\caption{Discovery image of {\name}. The top-left panel shows the ASAS-SN $V$-band reference image and the top-right panel shows the ASAS-SN subtracted image from 2014 January 25. The bottom-left panel shows the archival SDSS $g$-band image of the host galaxy and the bottom-right panel shows an LT 2-m $g$-band image from 2014 February 08. The dates of the observations are listed in each panel, and the lower panels show a smaller field of view, indicated by the red box in the top-left panel. The red circles have radii of 5\farcs{0} and are centered on the host position.}
\label{fig:disc}
\end{figure}

Here we describe the discovery and follow-up observations of {\name}, a potential TDE. The transient was discovered by ASAS-SN, a long-term project to monitor the whole sky on a rapid cadence to find nearby supernovae and other bright transients (see \citealt{shappee13} for details), such as AGN activity \citep[e.g.,][]{shappee13}, extreme stellar flares \citep[e.g.,][]{schmidt14}, outbursts in young stellar objects \citep[e.g.,][]{holoien14}, and cataclysmic variable stars \citep[e.g.,][]{stanek13,kato13}. Our transient source detection pipeline was triggered on 2014 January 25, detecting a new source with $V=17.1\pm0.1$~mag \citep{prieto14}. The object was also detected on 2014 January 26 at roughly the same magnitude, but is not detected ($V\ga18$~mag) in data obtained on 2014 January 1 and earlier. A search at the object's position in the Sloan Digital Sky Survey Data Release 9 \citep[SDSS DR9;][]{ahn12} catalog revealed the source of the outburst to be the inclined spiral galaxy {\galname} at redshift $z=0.0436$, corresponding to a luminosity distance of $d=193$~Mpc ($H_0=70$~km~s$^{-1}$~Mpc$^{-1}$, $\Omega_M=0.3$, $\Omega_{\Lambda}=0.7$), and that the ASAS-SN source position was consistent with the center of the host galaxy. Follow-up images obtained on 2014 January 27 with the Las Cumbres Observatory Global Telescope Network (LCOGT) 1-m telescope at McDonald Observatory \citep{brown13}, the 2-m Liverpool Telescope (LT) \citep{steele04}, and the {\swift} UltraViolet and Optical Telescope \citep[UVOT;][]{roming05} confirmed the detection of the transient. After astrometrically aligning an LT image of the source in outburst with the archival SDSS image of the host galaxy, we measure an offset of $0.28\pm0.45$ pixels ($0.09\pm0.14$~arcseconds) between the position of the brightest pixel in the host galaxy in the LT image and the position of the brightest pixel in the SDSS image. This indicates that the source of the new flux is consistent with the center of the galaxy, providing evidence for a TDE interpretation. Figure~\ref{fig:disc} shows the ASAS-SN $V$-band reference and subtracted images of the source as well as SDSS pre-discovery and LT $g$-band images.

The archival SDSS spectrum of the host is that of an early-type spiral with little evidence of emission lines from an AGN, although it does show [\ion{O}{3}]~5007 in emission indicating that there is some recent star-formation. A transient classification spectrum obtained on 2014 January 29 with the Dual-Imaging Spectrograph (DIS) mounted on the Apache Point Observatory (APO) 3.5-m telescope showed a blue continuum as well as a broad ($\rm FWHM \simeq 17000$~km/s) H$\alpha$ line. The blue continuum and H$\alpha$ emission suggested that this transient was likely a young Type~II SN, but the proximity to the galactic nucleus and its absolute magnitude at discovery ($M_V \sim -19.3$~mag from the ASAS-SN host-subtracted image) made a tidal capture event a potential alternative. We decided to start a follow-up campaign in order to fully characterize this interesting transient.

In \S\ref{sec:obs} we describe pre-outburst archival observations, including both photometry and spectroscopy of the host galaxy, as well as new data taken of the transient during our follow-up campaign. In \S\ref{sec:analysis} we analyze these data and describe the properties of the transient. Finally, in \S\ref{sec:disc} we compare these properties to those of supernovae, AGN, and other proposed TDEs to examine the nature of the object.

\section{Observations and Survey Data} 
\label{sec:obs} 

In this section we summarize the available archival data of the transient host galaxy as well as our new photometric and spectroscopic observations of {\name}.

\subsection{Archival Photometry and Spectroscopy} 
\label{sec:archival}

We retrieved archival reduced images in $ugriz$ of {\galname} from SDSS DR9. We then measured the fluxes in a 5\farcs{0} aperture radius (the same aperture used to measure the source in follow-up data, chosen to match the {\swift} PSF and to minimize the effects of seeing variations on the photometry) to use for galaxy SED modeling and for subtracting the host galaxy fluxes from the transient fluxes. We also retrieved near-IR $JHK_s$ images from the Two-Micron All Sky Survey \citep[2MASS;][]{skrutskie06} and measured aperture magnitudes of the host galaxy in the same fashion. The measured SDSS and 2MASS magnitudes of the host galaxy are presented in Table~\ref{table:host_mags}. 

\begin{table}
\caption{Photometry of Host Galaxy}
\label{table:host_mags}
\begin{tabular}{@{}ccc}
\hline
Filter & Magnitude & Magnitude Uncertainty \\
\hline
$u$ & 19.16 & 0.03 \\
$g$ & 17.60 & 0.02 \\
$r$ & 16.94 & 0.02 \\
$i$ & 16.65 & 0.02 \\
$z$ & 16.45 & 0.02 \\
$J$ & 15.34 & 0.05 \\
$H$ & 14.73 & 0.10 \\
$K_s$ & 14.34 & 0.10 \\
\hline
\end{tabular}

\medskip
These are 5\farcs{0} radius aperture magnitudes from SDSS and 2MASS.
\end{table}

There are no archival Spitzer, Herschel, Hubble Space Telescope (HST), Chandra, or X-ray Multi-Mirror Mission (XMM-Newton) observations of the source. The host galaxy is not detected in the ROSAT All-Sky Survey with an upper flux limit of $3\times10^{-13}$~erg~s$^{-1}$~cm$^{-2}$ in the $0.1-2.4$~keV band \citep{voges99}, providing further evidence that the galaxy is inactive. We also retrieved archival mid-IR photometry from the Wide-field Infrared Survey Explorer \citep[WISE;][]{wright10}. From the WISE $W1$ and $W2$ measurements we calculate that the host galaxy has $(W1-W2)\simeq0.06\pm0.06$~mag, and this blue mid-IR color is further evidence against AGN activity \citep[e.g.,][]{assef13}.

We used the code for Fitting and Assessment of Synthetic Templates \citep[FAST v1.0;][]{kriek09} to fit stellar population synthesis (SPS) models to the 5\farcs{0} SDSS $ugriz$ and 2MASS $JHK_s$ magnitudes of the host galaxy. The fit was made assuming a CCM extinction law \citep{cardelli88} with $R_V=3.1$, an exponentially declining star-formation history, a Salpeter IMF, and the \citet{bruzual03} models. We obtained a good SPS fit (reduced $\chi_{\nu}^{2}=0.4$), with the following parameters: $A_V = 0.15_{-0.15}^{+0.15}$~mag, $M_{*}=(6.3_{-0.8}^{+0.6})\times10^{9}$~{\msun}, age$=2.2_{-0.2}^{+0.6}$~Gyr, and SFR$=(3_{-1}^{+5})\times 10^{-2}$~{\msun}~yr$^{-1}$. \edit{We note that these properties do not appear to be consistent with {\galname} being an E$+$A galaxy, as many of the hosts of the TDE candidates in \citet{arcavi14} were.} The FAST estimate of $A_V=0.15$~mag incorporates both Galactic and host extinction and this value is consistent with the Galactic extinction ($A_V=0.057$~mag based on \citealt{schlafly11}). In fits to the transient SED, we find no evidence for additional extinction, even though the {\swift} UV data, particularly the $UVM2$ band which lies on top of the 2200~\AA~extinction curve feature, is a powerful probe for additional dust \edit{(though this depends on the strength of the UV bump in the dust law)}. In the analyses of the event's SED which follow we only correct for this Galactic extinction.

We also obtained the spectrum of {\galname} from SDSS-DR9. The archival spectrum is dominated by absorption lines (e.g., Balmer lines, \ion{Ca}{1} G-band, \ion{Mg}{1}, \ion{Na}{1}, Ca H\&K, and the 4000~\AA~break) that are characteristic of intermediate-age and old stellar populations. This is consistent with the results of the FAST fit to the SED of the host. The spectrum does not show strong emission lines, except for the detection of an unresolved [\ion{O}{3}]~5007 line with $\rm FWHM \simeq 250$~km~s$^{-1}$ and integrated luminosity $\rm L_{[O\,III]}\simeq 2.4\times 10^{39}$~erg~s$^{-1}$. This is likely a sign of a low level of recent star formation, indicating the galaxy could host core-collapse supernova events, but without detecting other emission lines (e.g., H$\beta$, H$\alpha$, [\ion{N}{2}]) we cannot constrain the rate of star formation. We note that the [\ion{O}{3}]/H$\beta$ and [\ion{N}{2}]/H$\alpha$ ratios may indicate that the host contains a very weak Type 2 AGN, consistent with the analysis done in \citet{arcavi14}. However, other factors (e.g., the WISE photometry) argue against significant activity.

We use the $\rm FWHM\simeq250$~km~s$^{-1}$ of the [\ion{O}{3}] line from the latest spectroscopic epoch, presented in \S\ref{sec:spec}, to estimate a velocity dispersion for the galaxy and place an upper limit on the mass of the black hole of $M_{BH}<10^{6.8}$~{\msun} using the M-$\sigma$ relation from \citet{gultekin09}. SDSS reports a velocity dispersion of $\sigma=42.9\pm7.3$~km~s$^{-1}$, which is well below the velocity resolution of the SDSS spectrograph of $\sim 100$~km~s$^{-1}$, so we will conservatively regard the SDSS estimate as a limit of $\sigma<100$~km~s$^{-1}$. Using the same M-$\sigma$ relation from \citet{gultekin09} and the SDSS resolution of 100~km~s$^{-1}$ gives an upper limit of $M_{BH}\la10^{6.9}$~{\msun}, consistent with the limit we derive from the width of the [\ion{O}{3}] line. Finally, from the FAST fit, we have $M_{*}\sim10^{9.8}$~{\msun}, which is consistent with a bulge mass of $M_B\sim10^{9.4}$~{\msun} \citep{mendel14}. Using the $M_{B}$-$M_{BH}$ relation from \citet{mcconnell13} gives $M_{BH}\sim10^{6.8}$~{\msun}, which is again consistent with the limits derived from the host and transient spectra.

\subsection{New Photometric Observations}
\label{sec:phot}

\begin{figure}
\centering
\includegraphics[width=0.95\linewidth]{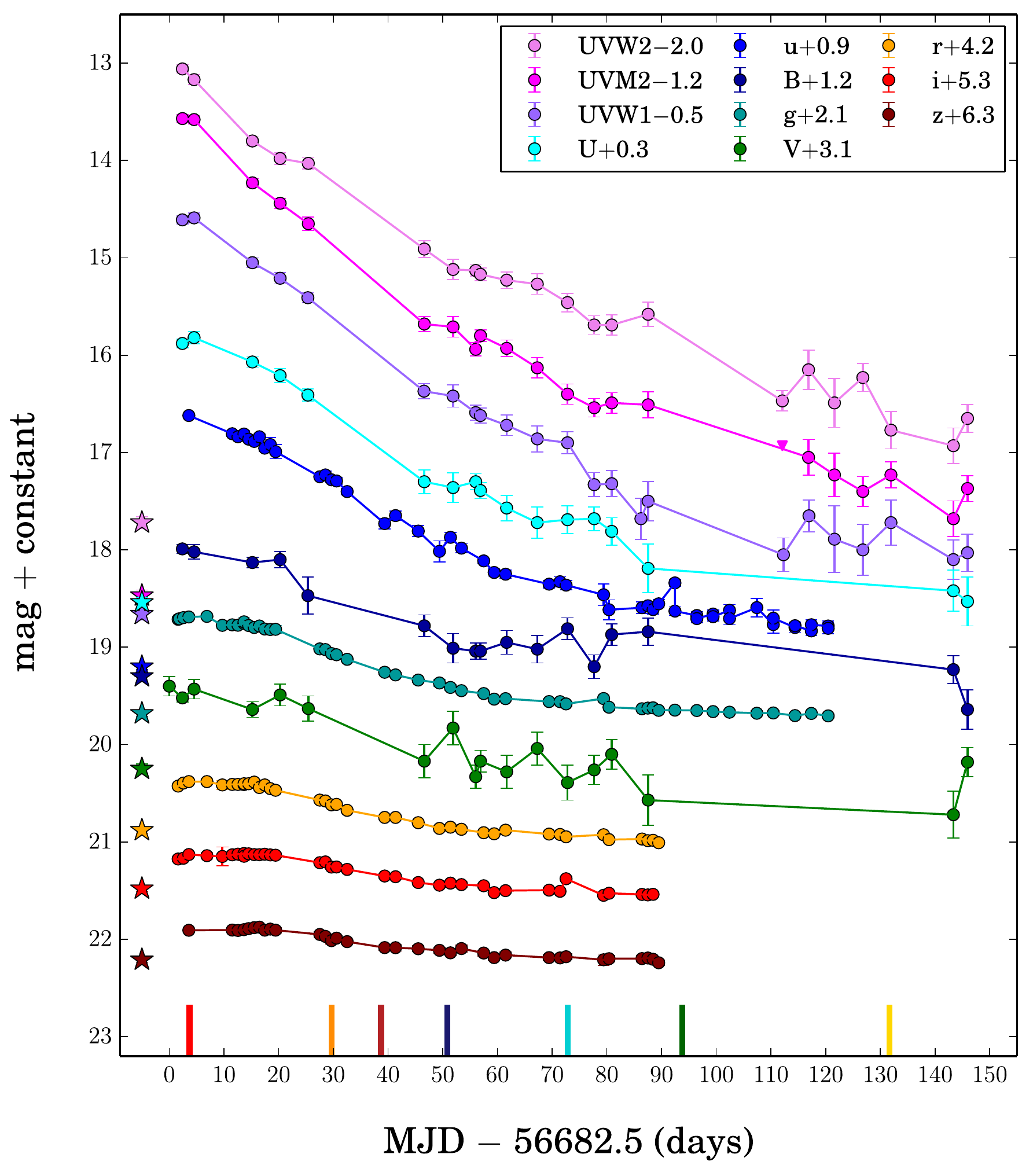} 
\caption{Light curves of {\name}, starting at discovery (MJD $=56682.5$) and spanning 146 days. Follow-up data obtained from {\swift} (UV $+$ optical), the LCOGT 1-m (optical), and the LT 2-m (optical) are shown as circles. 3-sigma upper limits are shown as triangles for cases where the source is not detected. All magnitudes are shown in the Vega system. The data are not corrected for extinction and error bars are shown for all points, but in some cases they are smaller than the data points. Host galaxy magnitudes measured by SDSS in a 5\farcs{0} aperture for $ugriz$ and synthesized from our host SED model for the {\swift} UVOT bands are shown as stars at $-5$ days. Dates of spectroscopic follow-up are indicated with vertical bars at the bottom of the figure with colors matching the corresponding spectra in Figures~\ref{fig:specevol} and~\ref{fig:specsub}. {\name} brightened by nearly 5 magnitudes in the UV with respect to the host galaxy while brightening by a progressively smaller amount in redder filters. The relative decline rate with respect to the host is steepest for those filters with larger increases in brightness, with the {\swift} $UVM2$ magnitude declining at a rate of roughly 3.5 magnitudes per 100 days. Table~\ref{table:phot} contains all the follow-up photometric data.}
\label{fig:lightcurve}
\end{figure}

After detection of the transient, we were granted a series of {\swift} X-ray Telescope \citep[XRT;][]{burrows05} and UVOT target-of-opportunity (ToO) observations. The {\swift} UVOT observations of {\name} were obtained in 6 filters: $V$ (5468~\AA), $B$ (4392~\AA), $U$ (3465~\AA), $UVW1$ (2600~\AA), $UVM2$ (2246~\AA), and $UVW2$ (1928~\AA) \citep{poole08}. We used the UVOT software task {\it uvotsource} to extract the source counts from a 5\farcs0 radius region and a sky region with a radius of $\sim$40\farcs0. The UVOT count rates were converted into magnitudes and fluxes based on the most recent UVOT calibration \citep{poole08, breeveld10}. The UVOT Vega magnitudes are shown along with other photometric data in Figure~\ref{fig:lightcurve}.

The XRT was operating in Photon Counting mode \citep{hill04} during our observations. The data from all epochs were reduced and combined with the software tasks {\it xrtpipeline} and {\it xselect} to obtain an image in the 0.3$-$10 keV range with a total exposure time of $\sim42,030$~s. We used a region with a radius of 20 pixels (47\farcs{1}) centered on the source position to extract source counts and a source-free region with a radius of 100 pixels (235\farcs{7}) for background counts. We do not detect X-ray emission from {\name} to a 3-sigma upper limit of $5.9\times10^{-4}$~counts s$^{-1}$. To convert this to a flux, we assume a power law spectrum with $\Gamma=2$ and Galactic \ion{H}{1} column density \citep{kalberla05}, yielding an upper limit of $\sim 2.9\times10^{-14}$~erg cm$^{-2}$ s$^{-1}$.  At the host distance of $d=193$~Mpc, this corresponds to an upper limit of $L_X\leq1.3\times10^{41}$~erg s$^{-1}$ ($3.4\times10^7$~{\lsun}) on the average X-ray luminosity. The constraints for the individual {\swift} epochs are on average $\sim10$ times weaker, and we only consider the combined X-ray limit.

In addition to the {\swift} observations, we obtained $gri$ images with the LCOGT 1-m at the MacDonald Observatory and $ugriz$ images with the LT 2-m telescope. We measured aperture photometry\footnote{We also attempted to do image subtraction with the SDSS archival images as templates. However, due to the lack of stars in the field-of-view close to {\name}, the quality of the subtractions was sub-optimal.} using a 5\farcs0 aperture radius to match the host galaxy and {\swift} UVOT measurements. The photometric zero points were determined using several SDSS stars in the field. These data are shown in Figure~\ref{fig:lightcurve}.

Figure~\ref{fig:lightcurve} shows the UV and optical light curves of {\name} from MJD 56682.5 (the epoch of first detection) through our latest epoch of observations on MJD 56828 (146 days after first detection) without extinction correction or host flux subtraction. Also shown are the SDSS $ugriz$ magnitudes and synthesized {\swift} UVOT magnitudes of the host galaxy extrapolated from the host SED fit. With the host flux included, the light curve shows {\name} brightened much more strongly in the blue and UV filters than in the red bands, with the largest increase in the {\swift} $UVM2$ band (2246~\AA), where it brightened by $\Delta m_{UVM2}\sim-4.9$. The brightness also appears to be declining at a faster rate with respect to the host in bluer filters. We further analyze this light curve and compare it to SNe and TDEs in the literature in \S\ref{sec:lcanal}.

\begin{figure}
\centering
\includegraphics[width=0.95\linewidth]{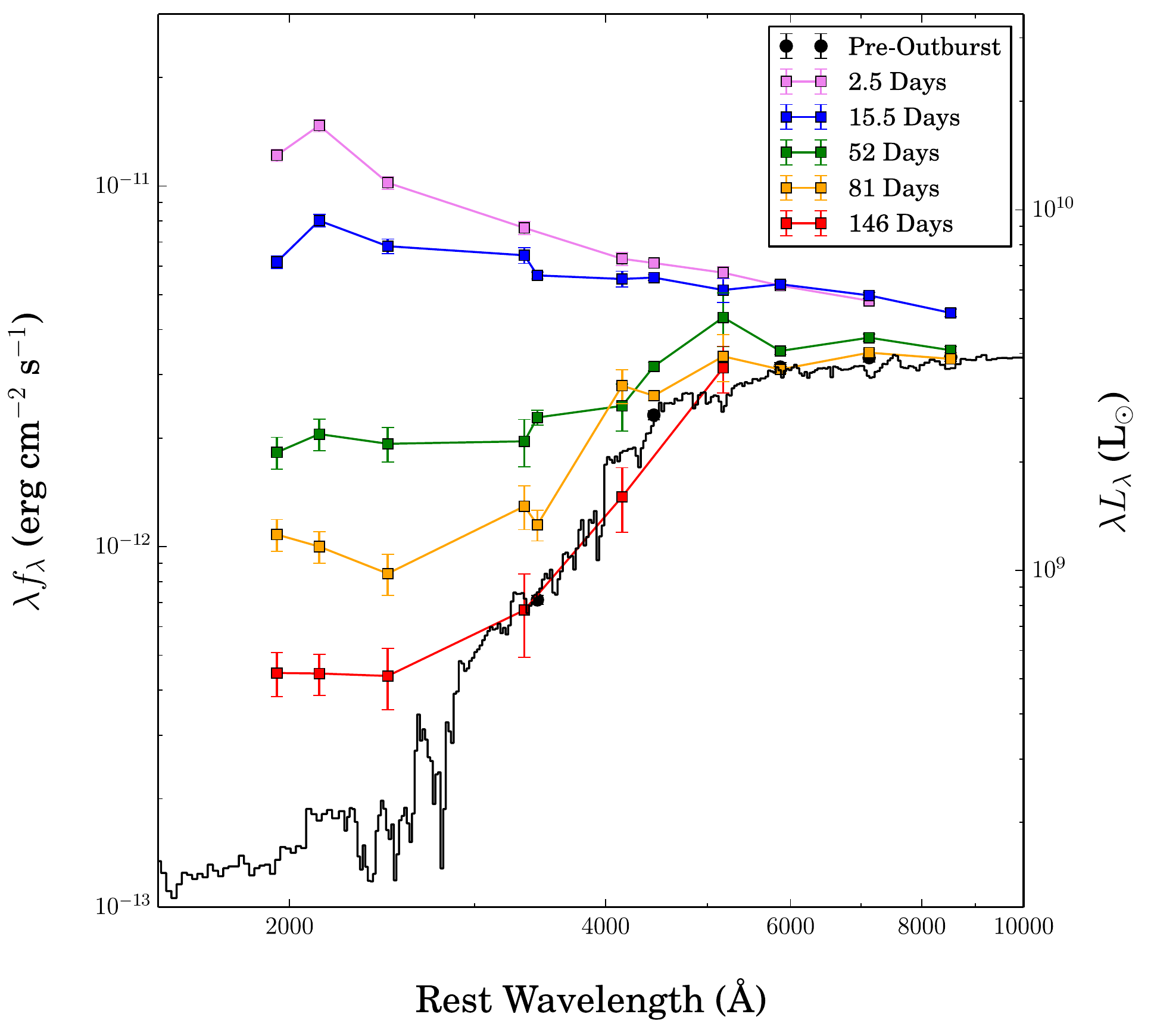}
\caption{Observed spectral energy distribution of {\name} and its host galaxy. The colored squares show the SED of {\name} at the different epochs noted in the legend (listed as days since discovery). The black circles show archival SDSS $ugriz$ data and the black line shows the best-fit host galaxy SED from FAST. All fluxes have been corrected for Galactic extinction and all data points include error bars, although they can be smaller than the data points.}
\label{fig:sed}
\end{figure}

After correcting our photometric measurements for Galactic extinction, we construct SEDs for 5 follow-up epochs. These are shown with the extinction-corrected SDSS archival data and host SED fit from FAST in Figure~\ref{fig:sed}. 

\subsection{New Spectroscopic Observations}
\label{sec:spec}

\begin{figure}
\centering
\includegraphics[width=0.9\linewidth]{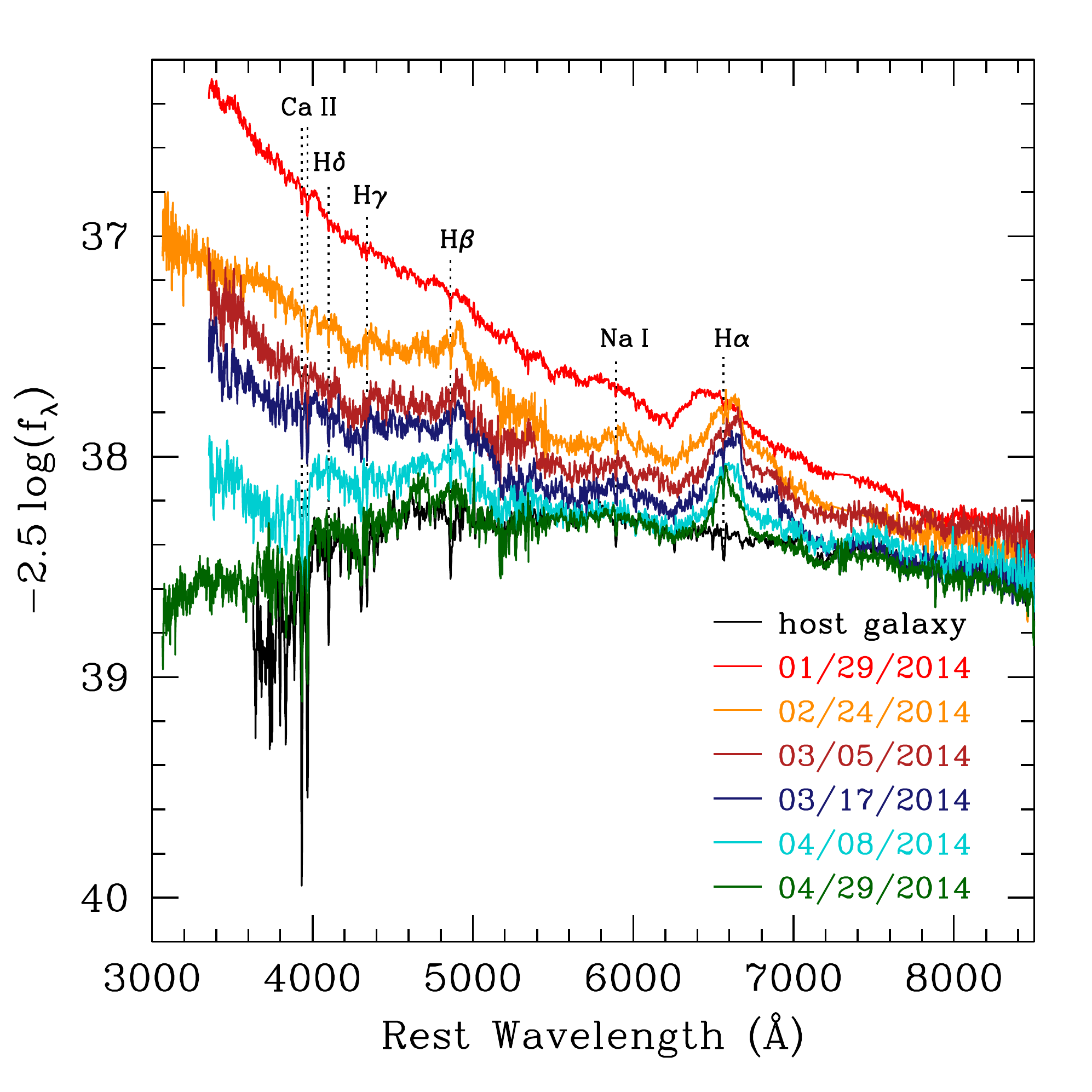}
\caption{Spectral time-sequence of {\name} during the outburst. Each spectrum shows the UT date it was obtained. Also plotted is the archival SDSS host spectrum, in black. Absorption features from the host galaxy are identified with black dotted lines. The transient spectra continue to show prominent broad H$\alpha$ emission at all epochs, as well as other Balmer lines.}
\label{fig:specevol}
\end{figure}

\begin{figure}
\centering
\includegraphics[width=0.9\linewidth]{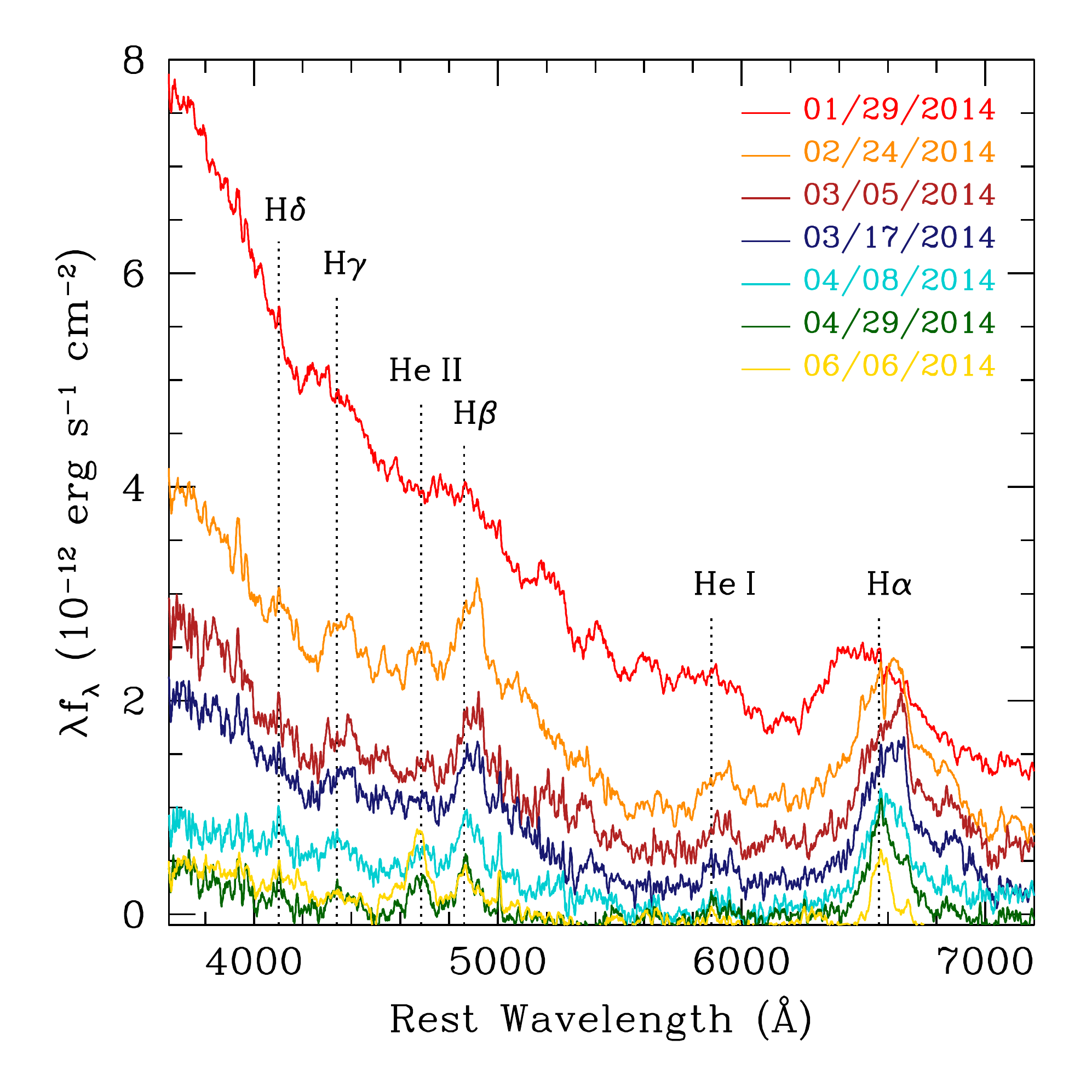}
\caption{Host-galaxy-subtracted spectral time-sequence of {\name}. \edit{Each spectrum shows the UT date it was obtained.} Prominent emission features are identified with black dotted lines. The transient spectra show many broad emission features in all epochs, and blue continuum emission is still present at wavelengths shorter than $\sim 4000$~{\AA} in the latest spectrum from 2014 June 6. In later epochs, the \ion{He}{2}~4686 line has become stronger relative to the Balmer lines}
\label{fig:specsub}
\end{figure}

We obtained seven low- and medium-resolution optical spectra of ASASSN-14ae spanning more than four months between 2014 January 29 and 2014 June 6. The spectra were obtained with DIS mounted on the Apache Point Observatory 3.5-m telescope (range $3500-9800$~\AA, $\rm R\sim 1000$) and with the Multi-Object Double Spectrographs (MODS; \citealt{Pogge2010}) on the 8.4-m Large Binocular Telescope (LBT) on Mount Graham (range $3200-10000$~\AA, $\rm R\sim 2000$). The spectra from DIS were reduced using standard techniques in IRAF and the spectra from MODS were reduced using a custom pipeline written in IDL\footnote{\url{http://www.astronomy.ohio-state.edu/MODS/Software/modsIDL/}}. We applied telluric corrections to all the spectra using the spectrum of the spectrophotometric standard observed the same night. We calculated synthetic $r$-band magnitudes and scaled the fluxes in each spectrum to match the $r$-band photometry. Figure~\ref{fig:specevol} shows a montage of the flux-calibrated spectra from both DIS and MODS, while Figure~\ref{fig:specsub} shows the same six spectra with the host galaxy spectrum subtracted.

The main characteristics of the spectra of {\name} are the blue continuum, consistent with the photometric measurements, and the detection of broad Balmer lines in emission, which are not present in the host galaxy spectrum. The H$\alpha$ line has $\rm FHWM \ga 8000$~km~s$^{-1}$ at all epochs and does not show a P-Cygni  absorption trough. The blue continuum  present in the first follow-up spectrum from 2014 January 29 becomes progressively weaker over time, with the spectrum from 2014 April 29 showing only slight emission above the host at wavelengths shorter than $\sim 4000$~\AA. While the latest spectrum from 2014 June 6 appears to show more blue continuum emission than the previous epoch, the flux calibration and host subtraction for this spectrum are uncertain, and the corresponding {\swift} photometry indicates the UV emission of the source should be fading. The broad H$\alpha$ emission feature becomes stronger relative to the continuum (higher equivalent width) after the initial spectrum and continues to show strong emission in all later epochs. Other broad emission features can be seen as well, including a \ion{He}{2}~4686 line which has become stronger in equivalent width relative to the Balmer lines in the latest spectrum. We further analyze the features of these spectra in comparison to SNe, AGN, and TDEs in \S\ref{sec:specanal}.

\section{Analysis}
\label{sec:analysis}

\subsection{Light Curve Analysis}
\label{sec:lcanal}

After correcting both the host and transient fluxes for Galactic extinction, we produced host-subtracted light curves for all 9 photometric filters. From these data we calculate peak absolute magnitudes and decline rates for all {\swift} filters, which are reported in Table~\ref{table:swiftmags}. Comparison with luminous supernovae SN 2008es \citep{miller09} and SN 2009kf \citep{botticella10}, both of which had absolute $V$-band magnitudes roughly equal to or greater than that of {\name}, shows that the UV decline rates of these highly luminous supernovae are much faster than what we observe for {\name}, indicating that a supernova explanation for the event is disfavored.

\begin{table*}
\begin{minipage}{145mm}
\caption{Peak absolute magnitudes and estimated decline rates of {\name} in {\swift} filters}
\label{table:swiftmags}
\begin{tabular}{@{}ccccc}
\hline
Filter & Absolute Magnitude & Magnitude Uncertainty & Decline Rate (mag/100 days) & Decline Rate Uncertainty \\
\hline
$V$ & $-19.5$ & 0.20 & 1.9 & 0.30  \\
$B$ & $-19.4$ & 0.07 & 3.6 & 0.36  \\
$U$ & $-19.8$ & 0.05 & 3.5 & 0.30 \\
$UVW1$ & $-19.8$ & 0.04 & 4.0 & 0.30 \\
$UVM2$ & $-20.0$ & 0.04 & 3.3 & 0.16 \\
$UVW2$ & $-19.7$ & 0.04 & 3.6 & 0.13 \\
\hline
\end{tabular}
\end{minipage}
\end{table*}

\begin{figure}
\centering
\includegraphics[width=0.95\linewidth]{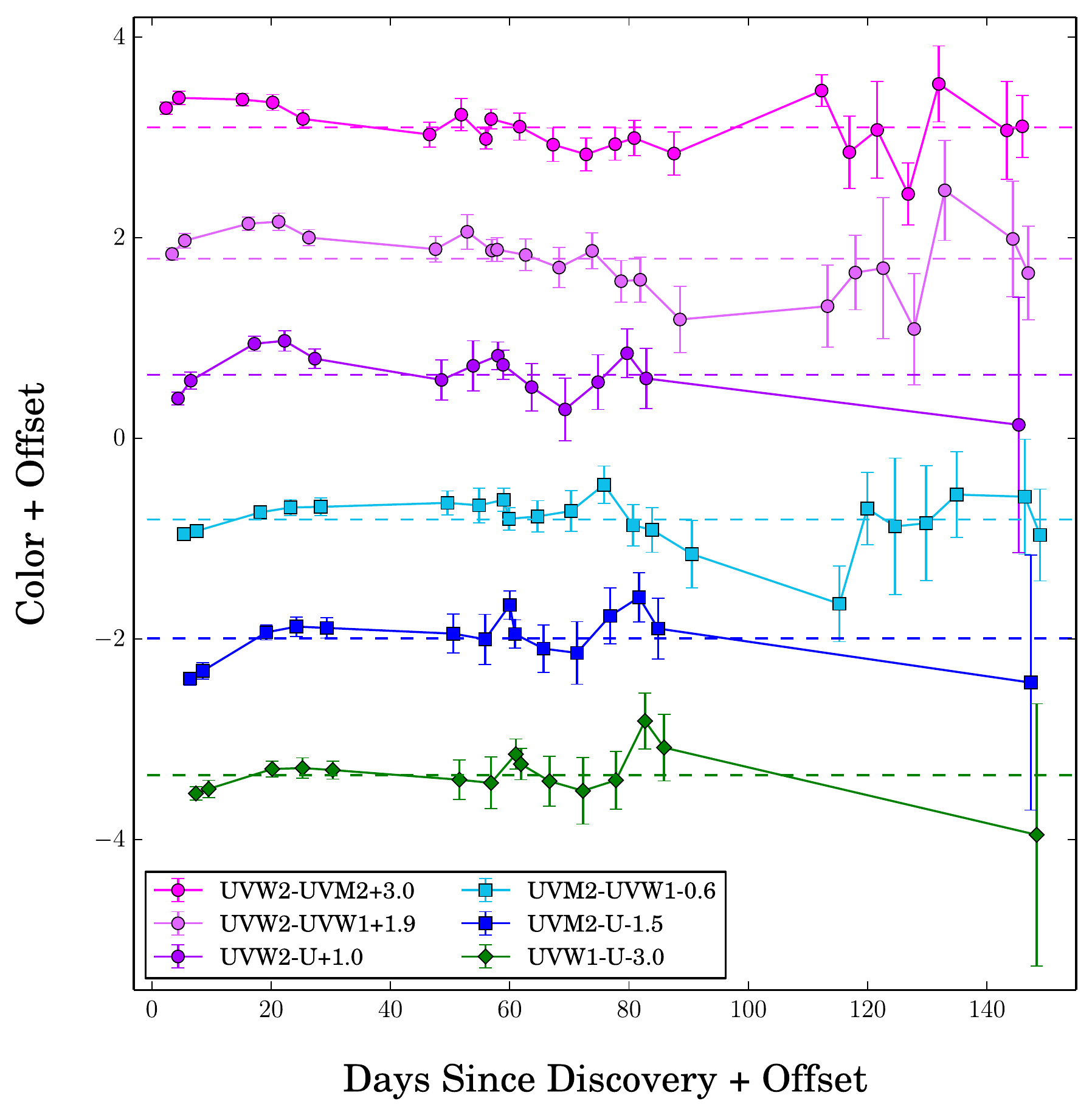}
\caption{$(UV-UV)$ and $(UV-U)$ color evolution of {\name} for all {\swift} UV bands. All fluxes used to calculate the colors shown were corrected for Galactic extinction and host-subtracted. $(UVW2-X)$ colors are shown as circles colored different shades of purple, $(UVM2-X)$ colors are shown as squares colored different shades of blue, and $(UVW1-U)$ is shown as diamonds and colored green. Each color term is offset in magnitude by a constant indicated in the legend and offset in epoch by 1 day from the term above it in order to make the plots easier to read. Horizontal dashed lines are centered on the average value of the color term plotted in the same color and are shown to aid the eye in seeing the general shape of the curves. {\name} becomes slightly bluer in $(UVW2-X)$ colors and slightly redder in $(UVW1-U)$, but all terms show only slight evolution over the time shown.}
\label{fig:color_evol}
\end{figure}

\begin{figure}
\centering
\includegraphics[width=0.95\linewidth]{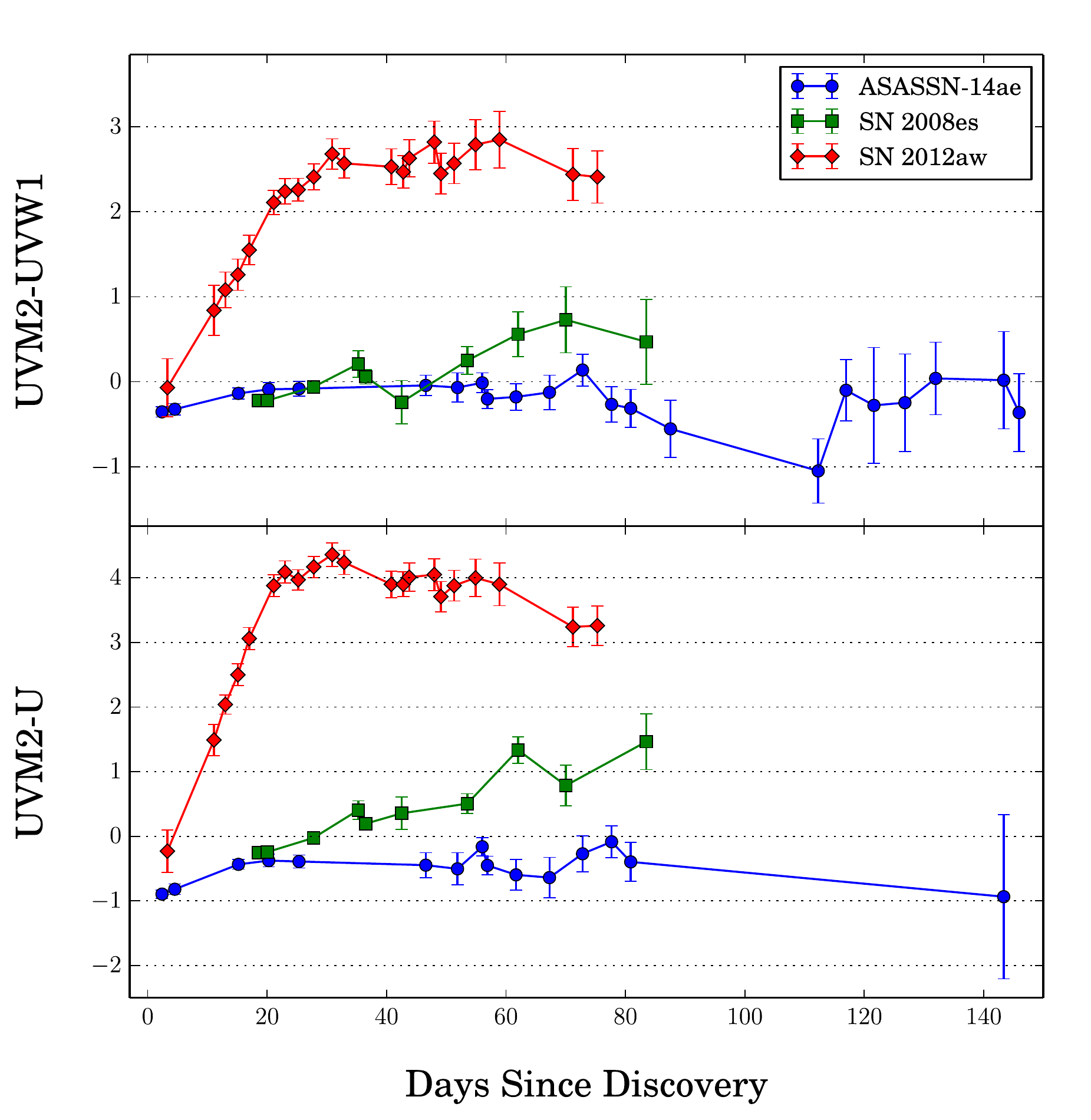}
\caption{Comparison of $(UVM2-UVW1)$ (top panel) and $(UVM2-U)$ (bottom panel) color evolution between {\name} (blue circles); SN 2008es, a super-luminous SN IIL \citep[][green squares]{gezari09}; and SN 2012aw, a SN IIP \citep[][red diamonds]{bayless13}. K-correction has been applied to the photometry for SN 2008es. {\name} shows little evolution in either color while SN 2008es becomes \edit{redder in both colors} and SN 2012aw becomes significantly redder over the first $\sim20$~days after detection and remains roughly constant thereafter.}
\label{fig:color_comp}
\end{figure}

{\name}'s UV-UV and UV-optical color evolution are also atypical of hydrogen-rich supernovae with broad lines. Figure~\ref{fig:color_evol} shows the full host-subtracted UV-UV and UV-U color evolution for {\name} in all {\swift} filters and Figure~\ref{fig:color_comp} compares the $(UVM2-UVW1)$ and $(UVM2-U)$ colors to those of SN~2008es, a super-luminous SN IIL \citep{miller09,gezari09}, and SN~2012aw, a normal SN IIP \citep{bayless13}, which were also heavily observed with {\swift}. For SN 2008es we applied cross-filter K-corrections to obtain the rest-frame colors assuming a blackbody with $T_{\rm eff}= 8000$~K \citep{miller09,gezari09}, but we did not apply these corrections to SN 2012aw or {\name} as they are much lower redshift. {\name} shows almost no change in $(UVM2-UVW1)$ color and became only slightly redder in $(UVM2-U)$ during the $\sim80$~days shown in the Figure. In contrast, SN 2008es \edit{became redder in both colors over time}, while SN 2012aw became significantly redder in both colors over the first $\sim20$~days after discovery and then remained roughly constant in later epochs. {\name} looks like neither of these, and all its $(UV-UV)$ and $(UV-U)$ colors show little change over the time shown in Figure~\ref{fig:color_evol}, implying the most likely SN types that could produce the observed spectra of {\name} are unlikely to be the sources of the transient.

\subsection{SED Analysis}
\label{sec:sedanal}

\begin{figure}
\centering
\includegraphics[width=0.95\linewidth]{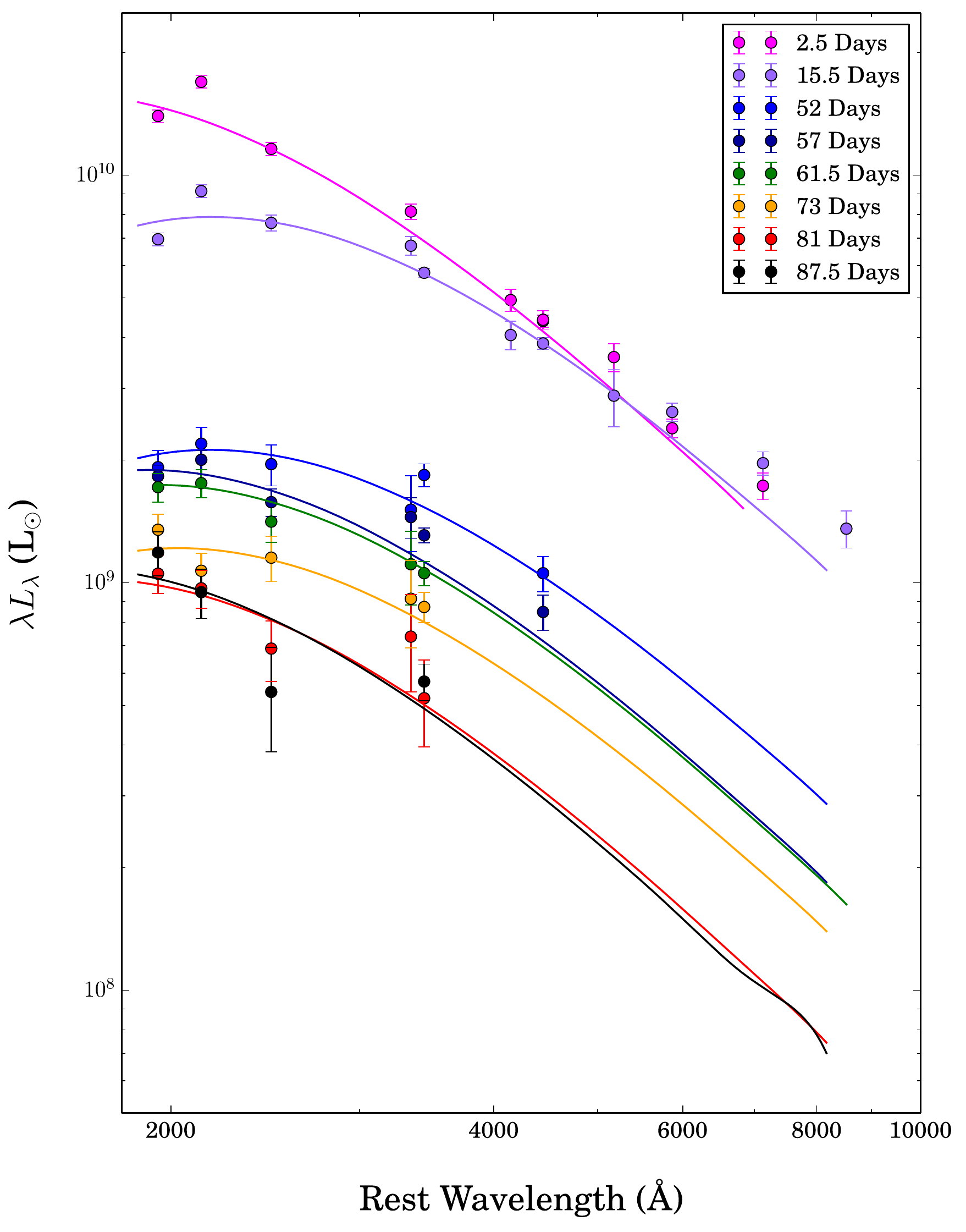}
\caption{Evolution of the SED of {\name} (shown in different colors) along with the best-fitting blackbody models for each epoch. Only epochs with both {\swift} and ground data taken within 0.5 days of each other and only data points with $f_{sub}/f_{host}\geq0.3$ are shown. All data points have been extinction-corrected and include error bars, although they can be smaller than the data points. At early epochs, the blackbody fits are not able to replicate the apparent excess in the {\swift} $UVM2$ band.} 
\label{fig:sed_fit}
\end{figure}

\begin{figure*}
\begin{minipage}{175mm}
\centering
\subfloat{{\includegraphics[width=0.45\textwidth]{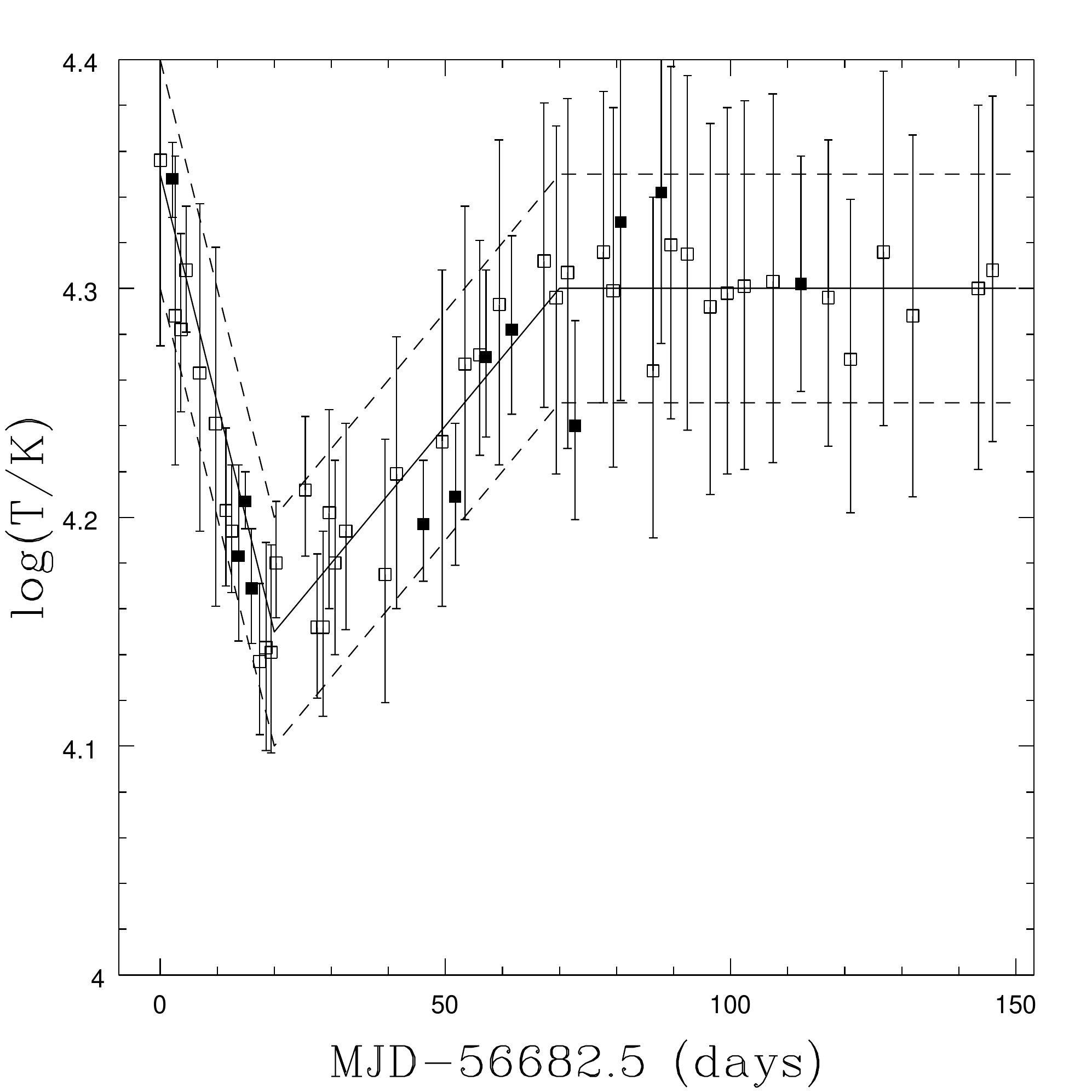}}}
\subfloat{{\includegraphics[width=0.45\textwidth]{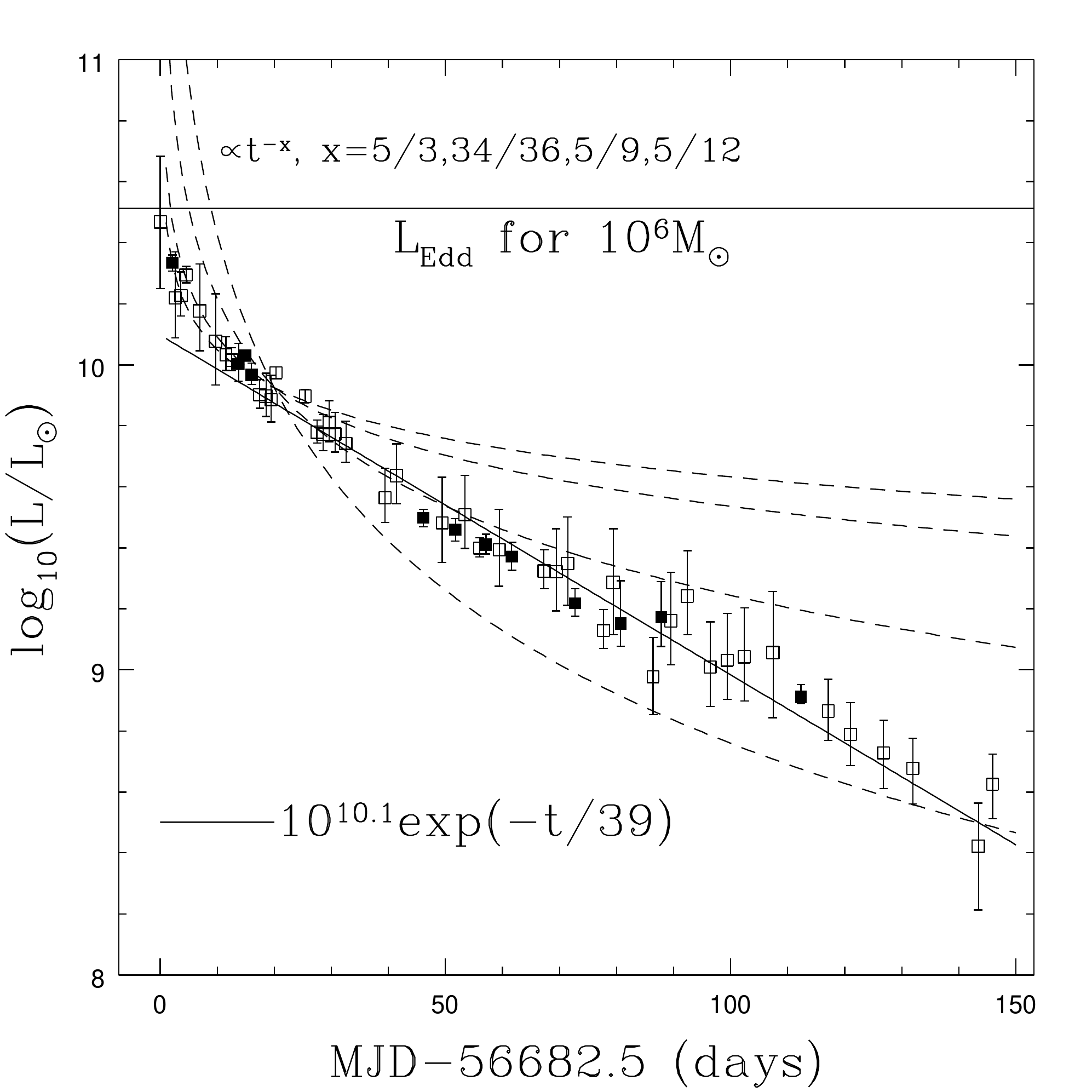}}}
\caption{\emph{Left Panel}: Evolution of \name's blackbody temperature with temperatures fit with a prior (open points) and without a prior (filled points). The horizontal lines show our temperature prior, with the solid line showing our central temperature prior and the dashed lines showing the 1$\sigma$ spread in the prior. The temperature of the source falls from $T\sim20,000$~K to $T\sim15,000$~K during the first $\sim20$~days of the outburst, then rises again to $T\sim20,000$~K over the next $\sim50$~days before remaining roughly constant for the rest of the period shown. \emph{Right Panel}: Evolution of \name's luminosity over time. Dashed lines show popular power law fits for TDE luminosity curves $L\propto t^{-x}$ \citep[e.g.,][]{strubbe09,lodato11} while the diagonal solid line shows an exponential fit. The solid horizontal line shows the Eddington luminosity for a $M=10^6$~{\msun} black hole. The exponential model appears to fit the luminosity curve of {\name} better than any of the power law fits typically used for TDEs.} 
\label{fig:temp_lum}
\end{minipage}
\end{figure*}

Using the host-subtracted fluxes of {\name} we fit the transient SEDs with blackbody curves using Markov Chain Monte Carlo (MCMC) methods. The evolution of the source's SED along with the best-fit blackbody curves are shown in Figure~\ref{fig:sed_fit}. At early epochs, the blackbody fit is not able to replicate the apparent excess in the $UVM2$ (2246~\AA) filter. This excess is not created by the extinction correction and corresponds to no obvious emission line. Using the best-fit blackbody curves we estimate the temperature and luminosity evolution of {\name}. The derived estimates, along with 90\% confidence errors and the $\chi^2$ values of the best-fit blackbody curve, are given in Table~\ref{table:temp_lum} and shown in Figure~\ref{fig:temp_lum}. When there were {\swift} observations the temperature was estimated as a free parameter. When the UV data were not available, the temperature was constrained by a prior shown by the solid line in Figure~\ref{fig:temp_lum} that roughly tracks the epochs with UV data. There are some degeneracies between the temperature and luminosity because much of the luminosity is farther in the UV than our data cover for these temperatures, resulting in relatively large uncertainties in some cases. In general, the temperature of the source falls from $T\sim20,000$~K to $T\sim15,000$~K during the first $\sim20$~days of the outburst, then rises again to $T\sim20,000$~K over the next $\sim50$~days, and then remains roughly constant for the rest of the period shown. Conversely, the luminosity fades steadily over the 150-day period shown. Integrating over the luminosity curve using the epochs with directly estimated temperatures gives a value of $E\simeq1.7\times10^{50}$~ergs for the total energy radiated by {\name} during this time. This only requires accretion of $\Delta M \sim 10^{-3}\eta_{0.1}^{-1}$~{\msun} of mass, where $\eta=0.1\eta_{0.1}$ is the radiative efficiency.

\begin{table*}
\begin{minipage}{120mm}
\caption{{\name} blackbody Evolution}
\label{table:temp_lum}
\begin{tabular}{@{}ccccc}
\hline
MJD & Best-Fit $\chi^2$ & Luminosity ($10^{9}$~${\lsun}$) & Temperature ($10^4$~K) & Radius ($10^{14}$~cm) \\
\hline
56684.6 & \s{35.5} & $21.6\pm1.3$ & $2.2\pm0.1$ & $\s{7.0}\pm0.3$ \\
56697.4 & \s{37.2} & $10.7\pm0.4$ & $1.6\pm0.0$ & $\s{9.3}\pm0.5$ \\
56698.5 & \s{27.0} & $\s{9.3}\pm0.8$ & $1.5\pm0.1$ & $10.4\pm0.8$ \\
56728.6 & \s{9.8} & $\s{3.1}\pm0.2$ & $1.6\pm0.1$ & $\s{5.3}\pm0.6$ \\
56734.2 & 15.0 & $\s{2.9}\pm0.2$ & $1.6\pm0.1$ & $\s{4.8}\pm0.6$ \\
56739.6 & 17.2 & $\s{2.6}\pm0.2$ & $1.9\pm0.2$ & $\s{3.4}\pm0.5$ \\
56744.1 & \s{1.7} & $\s{2.3}\pm0.2$ & $1.9\pm0.2$ & $\s{3.1}\pm0.4$ \\
56755.2 & \s{32.1} & $\s{1.7}\pm0.2$ & $1.7\pm0.2$ & $\s{3.1}\pm0.5$ \\
56763.2 & \s{9.7} & $\s{1.4}\pm0.4$ & $2.1\pm0.5$ & $\s{1.9}\pm0.7$ \\
56770.3 & 18.2 & $\s{1.5}\pm0.4$ & $2.2\pm0.4$ & $\s{1.9}\pm0.4$ \\
56794.8 & \s{28.4} & $\s{0.8}\pm0.1$ & $2.0\pm0.2$ & $\s{1.7}\pm0.3$ \\
\hline
\end{tabular}

\medskip
The results are given only for epochs with {\swift} data, where the temperature can be estimated without a prior.
\end{minipage}
\end{table*}

After the first $\sim10$~days, the luminosity evolution is well fit as an exponential $L\propto e^{-t/t_0}$ with $t_0\simeq39$~days as shown in Figure~\ref{fig:temp_lum}. This differs from most TDE models where the luminosity evolution is described as a power law $t^{-x}$ with $x\simeq5/12 - 5/3$ \citep[e.g.,][]{strubbe09,lodato11}. However, this temperature and luminosity behavior would be highly unusual for a supernova, which typically exhibit a temperature that drops considerably within days of the explosion along with either a relatively constant luminosity \citep[Type IIP; e.g.,][]{botticella10} or a declining luminosity \citep[Type IIn, IIL, Ic; e.g.,][]{miller09,inserra13,graham14}. 

\begin{figure}
\centering
\includegraphics[width=0.9\linewidth]{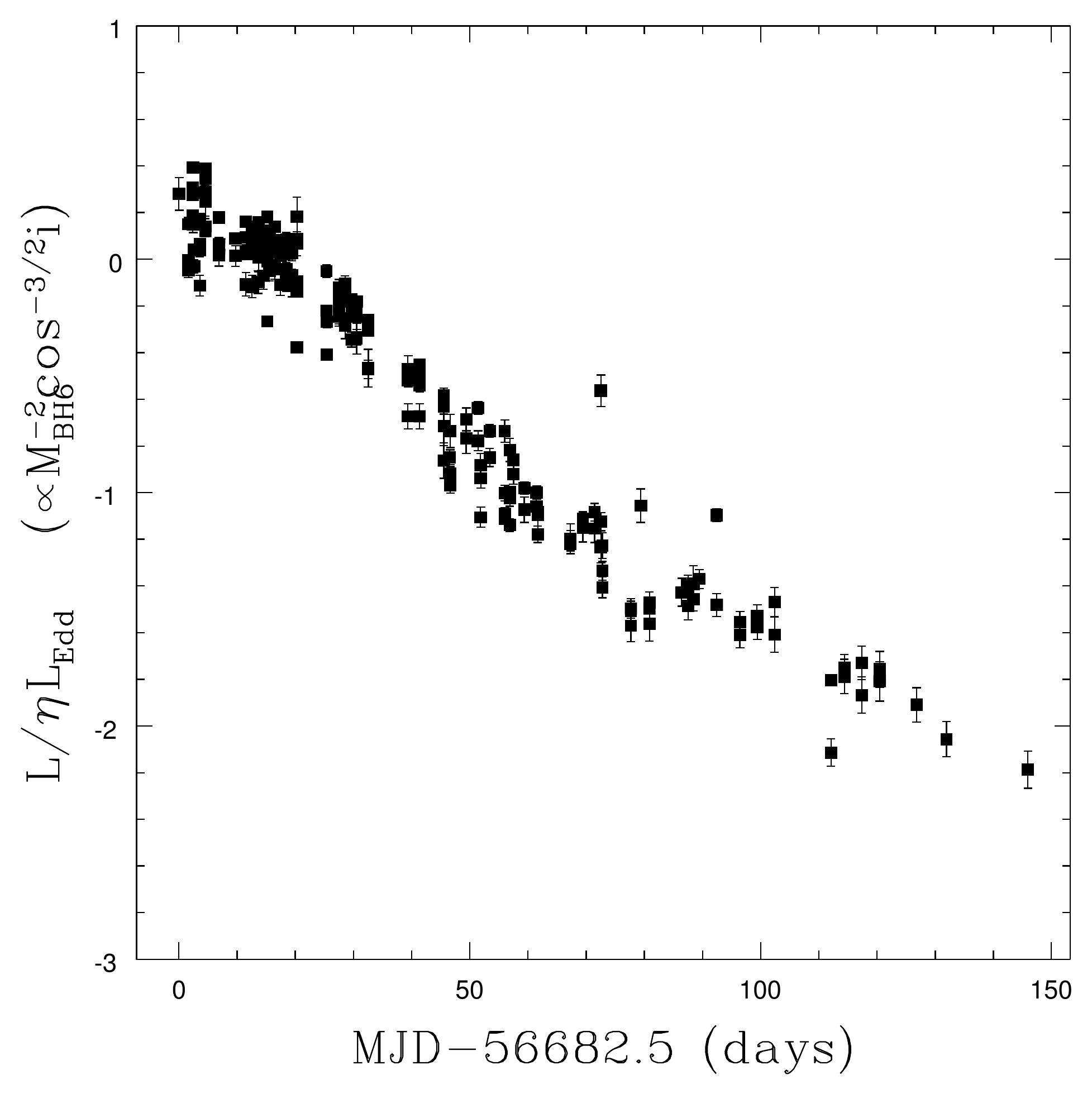}
\caption{Implied luminosity in Eddington units of {\name} using a thin disk model. The estimated $L/L_{Edd}$ depends on the black hole mass $M_{BH}=10^6 M_{BH6}$~{\msun}, the disk inclination factor $\cos i$ and the radiative efficiency $\eta = 0.1 \eta_{0.1}$. Raising the black hole mass scale to $M_{BH}\sim10^{6.5}$~{\msun} would produce $L/L_{Edd} \sim 1$ at peak. In general, the SED of the thin disk fits the data significantly worse than a blackbody (see Figure~\ref{fig:big_sed} below).}
\label{fig:thin_disk}
\end{figure}

While it is unlikely we are seeing direct emission from a thin disk, we can model the data using the surface brightness profile of a thin disk \citep{shakura73}.  We make the disk infinite, where the location of the inner edge is unimportant given our wavelength coverage.  Adding an outer edge could be used to make the profile rise more steeply towards shorter wavelengths.  Figure~\ref{fig:thin_disk} shows the implied luminosity of the disk in Eddington units, where the estimate of $L/L_{Edd}$ depends on the black hole mass $M_{BH}=10^6 M_{BH6}M_\odot$, the disk inclination factor $\cos i$ and the radiative efficiency $\eta = 0.1 \eta_{0.1}$ in the sense that raising the black hole mass, making the disk inclined to the line of site or lowering the radiative efficiency will reduce the observed luminosity relative to the Eddington limit.  In general, the SED of a thin disk fits the data significantly worse than a blackbody.  Like the bolometric luminosity, the estimated thin disk luminosity drops exponentially with time rather than as a power law, following a plateau at $L/L_{Edd} \simeq 10 \eta_{0.1} M_{BH6}^{-2} \cos^{-3/2} i$ for the first $\sim 20$~days.  Raising the black hole mass scale to $M_{BH}\sim10^{6.5}$~{\msun} would allow $L/L_{Edd} \sim 1$ at peak. The differences compared to the Eddington limit shown in Figure~\ref{fig:temp_lum} arise from the enormous increase in the unobserved hard UV emission which we discuss in \S\ref{sec:specanal} using the broad emission lines.

\subsection{Spectral Analysis}
\label{sec:specanal}

\begin{figure*}
\begin{minipage}{175mm}
\centering
\subfloat{{\includegraphics[width=0.45\textwidth]{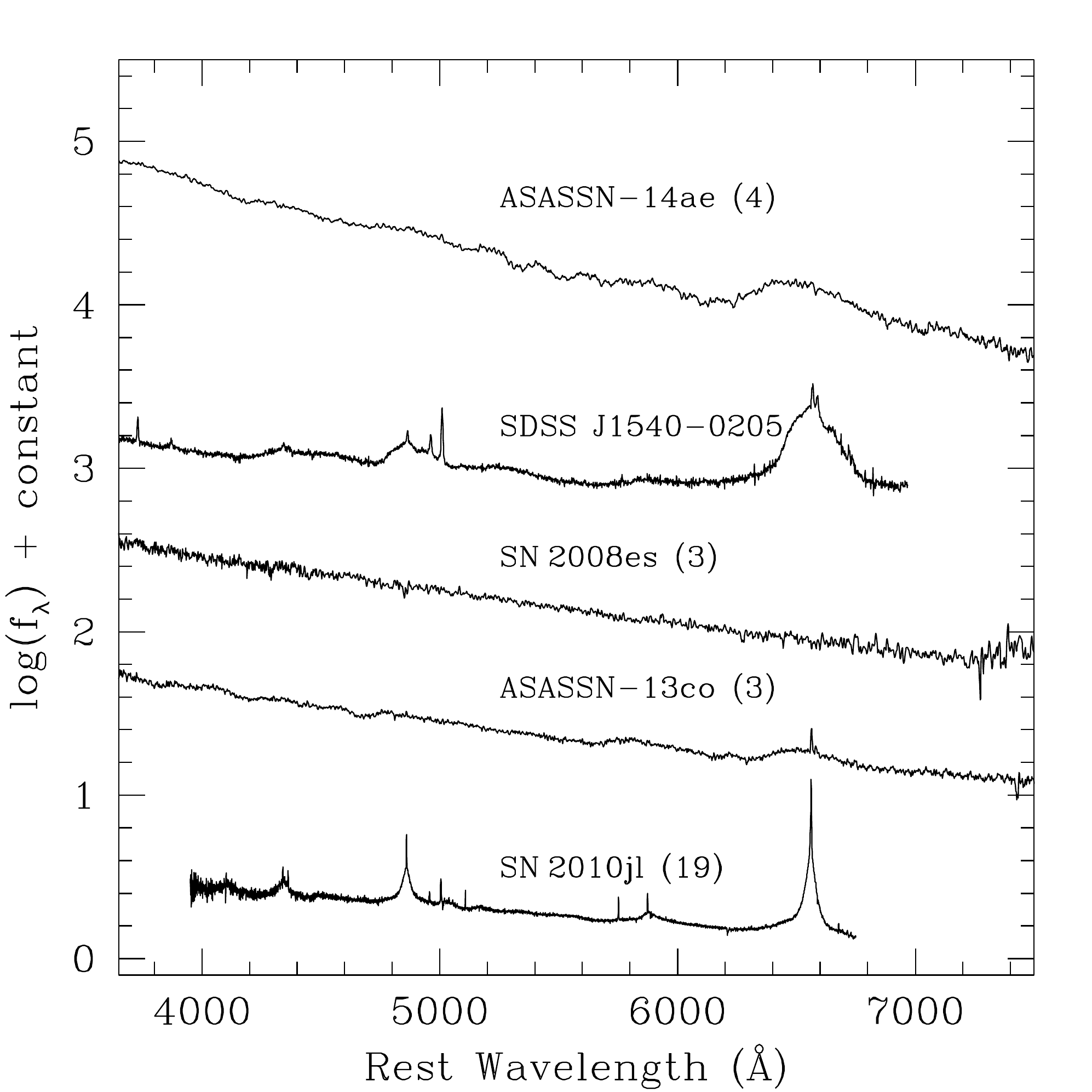}}}
\subfloat{{\includegraphics[width=0.45\textwidth]{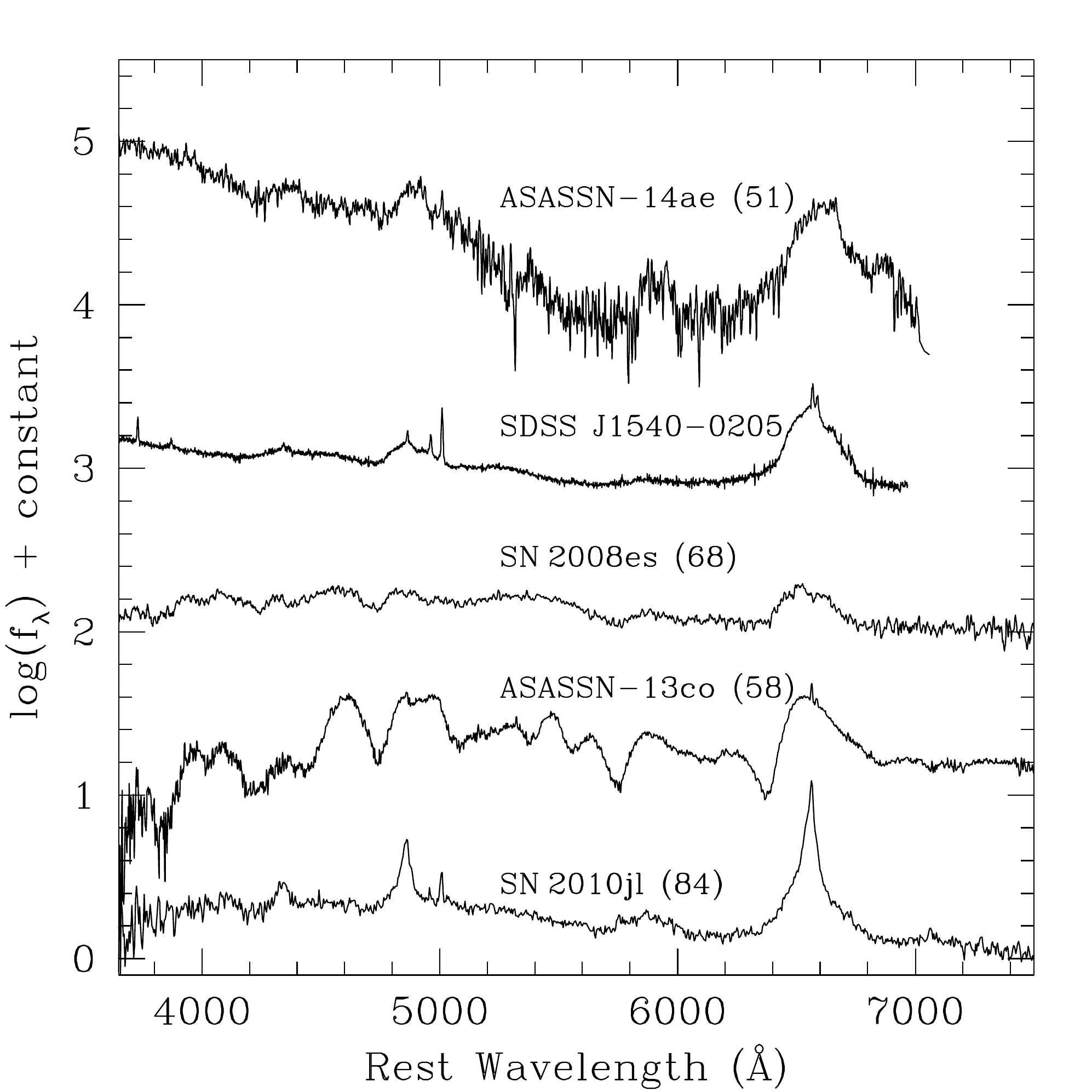}}}
\caption{Comparison of the host-subtracted spectra of {\name} with the spectra of the Type~IIn SN~2010jl \citep{stoll11,zhang12}, the Type~IIL SN~2008es \citep{miller09}, the Type~IIP ASASSN-13co \citep{holoien13}, and the broad-line AGN SDSS~J1540-0205 \citep{strateva03}. The left panel shows the spectra at an early phase and the right  panel at a later phase with respect to discovery or maximum light (except for the AGN). The days with respect to maximum light (SN~2008es, SN~2010jl) or discovery (ASASSN-14ae, ASASSN-13co) are shown in parenthesis, next to the names of the transients.} 
\label{fig:spec_comp}
\end{minipage}
\end{figure*}

The spectra of {\name} show broad Balmer lines in emission with a blue continuum, including an important contribution of the host galaxy (especially in the red) at late times (see Figure~\ref{fig:specevol}). We subtracted the SDSS host galaxy spectrum from {\name} spectra in order to compare to other objects and for analyzing the spectral line profiles. Figure~\ref{fig:spec_comp} shows a comparison of {\name} host-subtracted spectra at two different epochs after discovery with the spectra of SNe~II (SN~2010jl, ASASSN-13co, and SN~2008es) and a broad-line AGN (SDSS~J1540-0205). SN~2010jl is a luminous SN~IIn \citep[$M_V\simeq -20$~mag,][]{stoll11}, SN~2008es is a super-luminous SN~IIL \citep[$M_V\simeq -22$~mag,][]{miller09}, and ASASSN-13co is a luminous SN~IIP \citep[$M_V\simeq -18$~mag,][]{holoien13}. At the earliest epoch, the spectrum of {\name} is similar to ASASSN-13co and SN~2008es, dominated by a blue continuum. At the later epoch, the spectrum of {\name} stays blue, but the spectra of the SN~II ASASSN-13co and SN~2008es become significantly redder, consistent with the comparison in color evolution illustrated in Figure~\ref{fig:color_comp}. The spectral lines also show differences. In {\name} the H$\alpha$ is broad ($\rm FWHM \ga 8,000$~km~s$^{-1}$) at all times and does not show the P-Cygni profile that is characteristic of SNe~II with broad lines. The spectrum of {\name} is quite different from the Type~IIn SN~2010jl \citep{stoll11,zhang12} at early and late times, both in line profiles (SNe~IIn have narrower lines with $\rm FWHM \la 5,000$~km~s$^{-1}$) and continuum shape. The spectrum of the low-ionization broad-line AGN SDSS~J1540-0205 \citep{strateva03} does not resemble the spectrum of {\name} in the earlier epoch but shows interesting similarities with the spectrum of {\name} at $50$~days after discovery. In particular, it has a complex H$\alpha$ line profile, which is thought to be produced by emission from the accretion disk \citep{strateva03}. 

\begin{figure}
\centering
\includegraphics[width=0.9\linewidth]{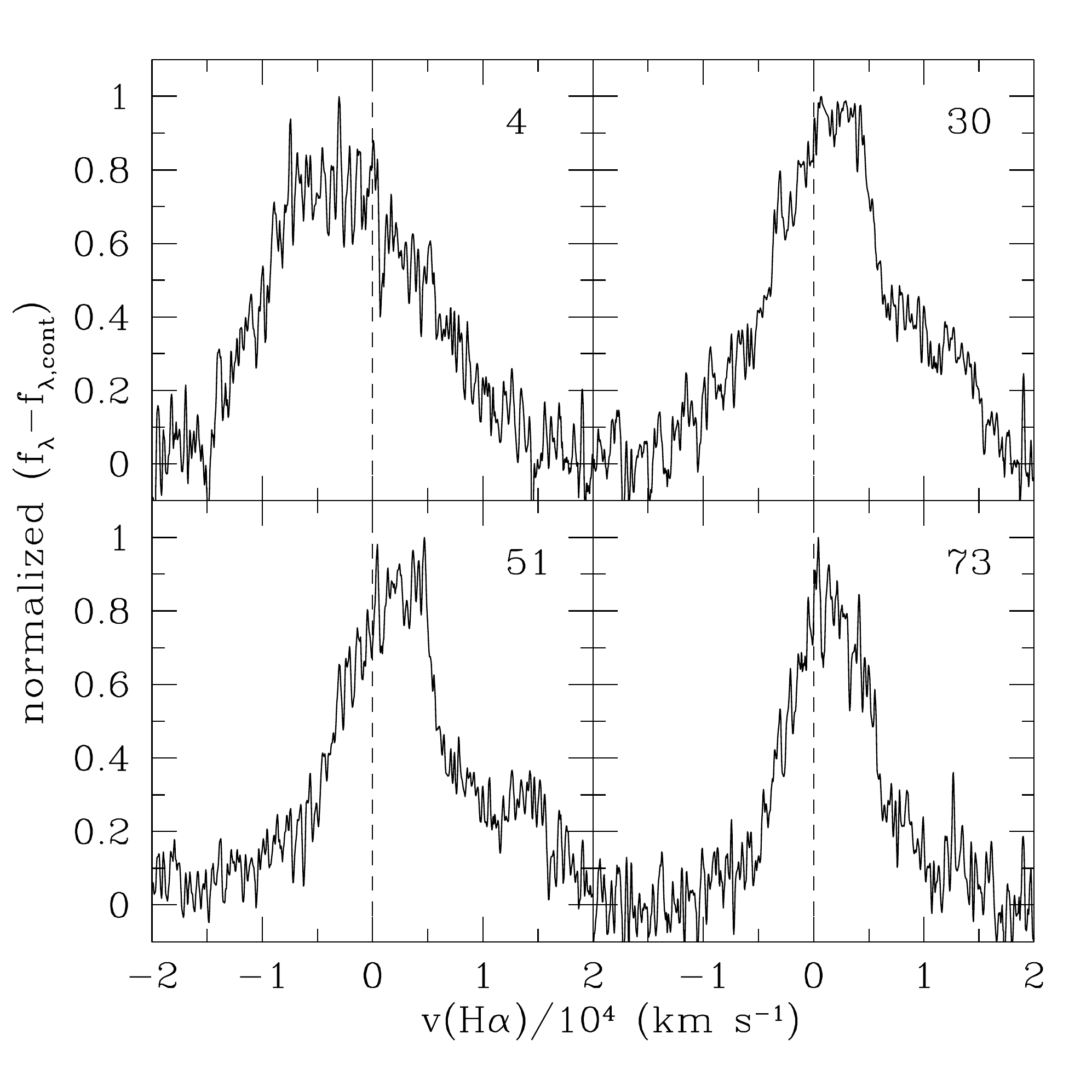}
\caption{Evolution of the H$\alpha$ line profile of {\name} as a function of time. We have subtracted the host galaxy spectrum and a low-order continuum defined locally around the line. The days since discovery are shown in the top-right part of each panel. The integrated luminosity of the H$\alpha$ line is $L_{H\alpha} \simeq 2.7\times10^{41}$~erg~s$^{-1}$ at four days, $L_{H\alpha} \simeq 3.3\times10^{41}$~erg~s$^{-1}$ at 30 days, $L_{H\alpha} \simeq 3.2\times10^{41}$~erg~s$^{-1}$ at 51 days, and $L_{H\alpha} \simeq 2.1\times10^{41}$~erg~s$^{-1}$ at 70 days.}
\label{fig:Halpha}
\end{figure}

In Figure~\ref{fig:Halpha} we show the evolution of the H$\alpha$ line profile of {\name} as a function of time. In the first epoch, 4~days after discovery, the line can be well-fit with a Gaussian profile centered at $\rm v_{peak}\sim -3,000$~km~s$^{-1}$ and with $\rm FWHM \simeq 17,000$~km~s$^{-1}$ and integrated luminosity $L_{H\alpha} \simeq 2.7\times10^{41}$~erg~s$^{-1}$. However, the line peak evolves to the red and the shape becomes significantly asymmetric in later epochs between $\sim 30-50$~days after discovery. The peak of the profile in these epochs is at $\rm v_{peak} \sim 3,000-4,000$~km~s$^{-1}$ and the blue/red wing of the line reaches $\sim -15,000/+20,000$~km~s$^{-1}$, showing a strong red asymmetry. At 70~days after discovery, the profile again becomes more symmetric and can be relatively well-fit using a gaussian with $\rm v_{peak}\sim +1,400$~km~s$^{-1}$ , $\rm FWHM \simeq 10,000$~km~s$^{-1}$ and integrated luminosity $L_{H\alpha} \simeq 2.1\times10^{41}$~erg~s$^{-1}$. In the spectrum from 2014 April 29, not shown in Figure~\ref{fig:Halpha}, the H$\alpha$ line has $\rm FWHM \simeq 8,000$~km~s$^{-1}$ and its integrated luminosity has only decreased by a factor of three since the first epoch, to $L_{H\alpha} \simeq 1.2 \times 10^{41}$~erg~s$^{-1}$. In the April 29 spectrum we also detect broad \ion{He}{2} 4686 with $\rm FWHM \simeq 6,000$~km~s$^{-1}$ and $L_{HeII} \simeq 3\times10^{40}$~erg~s$^{-1}$. In the latest spectrum from 2014 June 6, this \ion{He}{2} line has become stronger relative to the Balmer lines, but has the same FWHM.

In summary, the spectra of {\name} seem to be inconsistent with the spectra of SN~II and the H$\alpha$ emission line profile shows strong evolution during the event. Compared to the spectra of TDEs in the literature, the most similar to {\name} is SDSS TDE2 \citep{velzen11}, which showed a broad H$\alpha$ line with $\rm FWHM \simeq 8,000$~km~s$^{-1}$. The spectra of PS1-10jh \citep{gezari12b} showed a \ion{He}{2}~4686 line in emission with $\rm FWHM \simeq 9,000$~km~s$^{-1}$ and the spectra of PS1-11af \citep{chornock14} did not show any emission lines. The recent paper by \citet{arcavi14} presents spectra of multiple TDE candidates and shows that their spectra fall on a continuum, with some events being more He-rich and others being more H-rich. The spectra of {\name} resembles the other TDE candidates, and with both strong He and H emission, it appears to fall in the middle of the proposed continuum of spectral properties.

\begin{figure}
\centering
\includegraphics[width=0.9\linewidth]{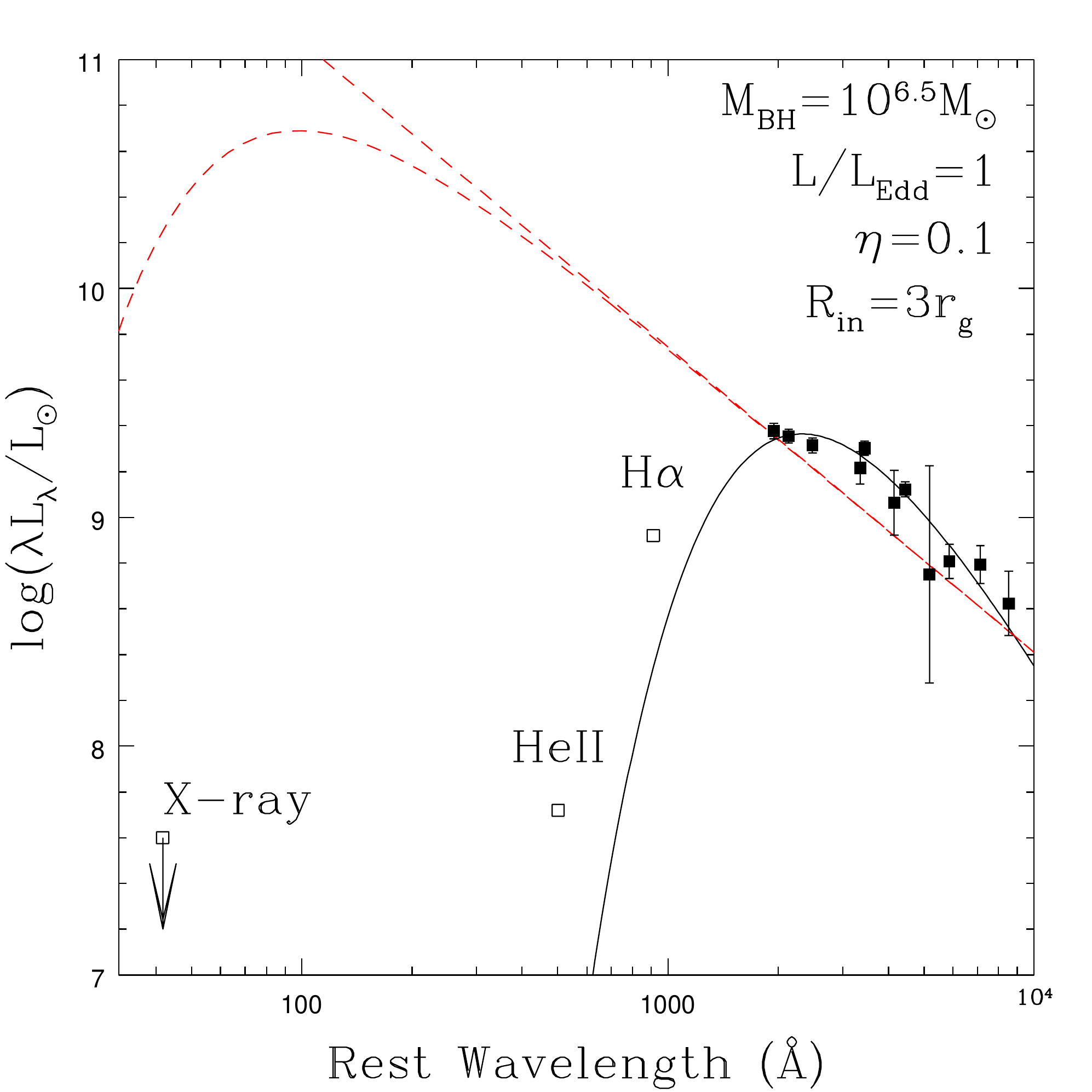}
\caption{SED of a thin disk with the black hole mass raised to $M_{BH}=10^{6.5}M_\odot$ so that a disk radiating at the Eddington luminosity is consistent with the observed UV emission from MJD $56729$ without an inner edge (straight red dashed line), and with an inner edge at $R_{in}=3r_g$ (curved red dashed line). Both models rise into the hard UV, producing far more ionizing flux than is required to produce the observed H$\alpha$ and \ion{He}{2} emission (unfilled boxes). The X-ray limit shown is based on {\swift} XRT data collected through 2014 April 24; including later data would make this limit tighter, as discussed in \S\ref{sec:phot}. These estimates of the ionizing luminosity can be shifted to higher luminosities as the inverse of the covering fraction, but H and He probably have to be shifted by similar amounts, which would imply that the spectrum must still be falling towards shorter wavelengths.}
\label{fig:big_sed}
\end{figure}

If we assume that the H$\alpha$ and \ion{He}{2} emission are driven by photoionization and recombination, we can gain some insight into the hard UV continuum and the physical conditions of the line emitting region.  In particular, if $\alpha_B$ and $\alpha_l$ are the case B recombination and line emission rate coefficients, and $E_i$ and $E_l$ are the energies of the ionization edge and the line, then we can estimate the luminosity at the ionization edge as $L_i = L_l (\alpha_B/\alpha_l)(E_i/E_l)$, which for $L_{H\alpha}\simeq2 \times 10^{41}$ and $L_{HeII} \simeq 3 \times 10^{40}$~erg~s$^{-1}$ implies ionizing luminosities of $3 \times 10^{42}$ and $ 2\times 10^{41}$~erg~s$^{-1}$, respectively, as shown in Figure~\ref{fig:big_sed}.  If we compare these estimates, we see that the SED probably requires some additional hard UV emission beyond that expected from the blackbody fits, but definitely has a sharp cutoff at wavelengths only somewhat shorter than the blackbody predictions. This assumes an emission line gas covering fraction $\epsilon \simeq 1$ near unity, and these ionizing luminosities can be shifted upwards as $\epsilon^{-1}$. However, the covering fractions for H and He are unlikely to differ enormously, so the SED likely must fall towards shorter wavelengths independent of the exact value of $\epsilon$. This is also consistent with the X-ray flux limit from \S\ref{sec:phot}.

For comparison, Figure~\ref{fig:big_sed} shows the SED of a thin disk, raising the black hole mass to $M_{BH}=10^{6.5}M_\odot$ so that an $L/L_{Edd} \simeq 1$ disk with efficiency $\eta=0.1$ is consistent with the observed UV emission near MJD $56729$ (as compared with Figure~\ref{fig:thin_disk}). The inner regions of the disk are much hotter than the blackbody, so the SED continues to rise into the hard UV, producing far more ionizing flux than is required.  This is true even when we add an inner edge at $R_{in}=3r_{g}$ where $r_g=GM_{BH}/c^2$ is the gravitational radius of the black hole.  Thus, while there must be some excess hard emission compared to a blackbody, it is likely less than for a thin disk.  Note that the effects of the inner disk edge only affect the very hard UV and X-ray emission, which is why we could ignore the inner edge in the SED models from \S\ref{sec:sedanal}.  The different shape of the thin disk model where we have the optical and near-UV SED also shows why the blackbody models are a better fit to the directly observed SEDs. Extrapolating the SED following the thin disk model would increase the total energy budget and accreted mass by roughly an order of magnitude.

We can also estimate the gas mass associated with the line emission if we assume that the line widths are related to orbital velocities. The emission radius for a velocity of $v=5000 v_5$~km/s is  of order
\begin{equation}
    r_\alpha \simeq 5 \times 10^{14} M_{BH6} v_5^{-2}~\hbox{cm}
\end{equation}
and the H$\alpha$ luminosity is of order
\begin{equation}
    L_{H\alpha} \simeq { 4 \pi \over 3 } r_\alpha^3 n^2 \alpha_l E_l
\end{equation}
where $\alpha_l$ and $E_l$ are the volumetric rate and line energy and we assume a spherical geometry.  This implies a characteristic number density of order
\begin{equation}
    n \simeq 3 \times 10^{10} L_{\alpha 41}^{1/2} v_5^3 M_{BH6}^{-3/2}~\hbox{cm}^{-3}
\end{equation}
which implies that the line emission should almost instantaneously track the hard UV emission because the recombination times are short.  The mass associated with the emission line region is then
\begin{equation}
   M_\alpha \simeq 0.016 L_{\alpha 41}^{1/2} M_{BH6}^{3/2} v_5^{-3} M_\odot
\end{equation}
which is an order of magnitude larger than the amount of accreted mass needed to power the transient.

\section{Discussion}
\label{sec:disc}

The transient {\name}, discovered by ASAS-SN on 2014 January 25, had a peak absolute $V$-band magnitude of $M_V\sim-19.5$ and position consistent to within \edit{$0.09\pm0.14$~arcseconds} of the center of {\galname}. However, it does not appear to be consistent with either a supernova or a normal AGN outburst. Its colors remain blue over 140 days since detection, rather than showing the rapid reddening seen in super-luminous SNe with similar absolute magnitudes, and {\name}'s temperature has remained relatively constant at $T\sim20,000$~K for the duration of the outburst while declining steadily in luminosity at a rate best fit by an exponential decay curve, behavior which is inconsistent with nearly all SNe. Finally, spectra of {\name} show a strong blue continuum and broad emission features, including Balmer lines and a \ion{He}{2} line, and its spectral evolution does not appear to match either those of SNe or AGN. While highly unusual AGN activity or a strange SN cannot be ruled out completely, the observational characteristics of {\name} disfavor both of these scenarios.

Archival photometry, spectroscopy, and SED fitting indicate that {\galname} appears to be an early-type galaxy with a generally intermediate-to-old-aged stellar population but some signs of recent star formation. While the recent star formation indicates that the galaxy could host a core-collapse SN, the supernova explanation is disfavored by photometric and spectroscopic observations, as previously mentioned. {\galname} shows spectral emission features indicating only a weak AGN host at best, and its mid-IR colors from WISE are inconsistent with significant AGN activity, further disfavoring normal AGN activity as an explanation for {\name}. 

Conversely, many of the observed properties of {\name} are consistent with other TDE candidates discussed in the literature. The blue colors, slow decline rate, and color evolution have been seen in many TDE candidates, and these transients are predicted to show a largely constant temperature and steadily declining luminosity curve. {\name}'s spectra also do not appear to be highly unusual for TDEs, and in fact are a very close match to spectra of the SDSS candidate TDE 2 \citep{velzen11} and the PTF candidate \edit{PTF09djl} \citep{arcavi14}. With both strong He and H emission lines, {\name} appears to fall in the middle of the He-rich-to-H-rich continuum proposed by \citet{arcavi14}.

Thus, we conclude that {\name} is probably a TDE. {\name} appears to be the lowest redshift candidate TDE discovered at optical or UV wavelengths, and continues to emit well above host galaxy levels in the UV over 140 days since discovery. In the most recent spectra, the optical continuum is again dominated by the host, but with a prominent, broad H$\alpha$ line and other, weaker Balmer and \ion{He}{2} lines.

The amount of mass associated with the event is small, roughly $10^{-3}$~{\msun} of accreted material is sufficient to power the transient, and $\sim10^{-2}$~{\msun} is associated with the line emission region. This suggests that this event is likely powered by tidally stripping the envelope of a giant rather than by the complete disruption of a main-sequence star, \edit{as described in \citet{macleod12}}, similar to the case seen with black hole candidate ESO 243-49 HLX-1 \citep{webb14}. The duration of a TDE can be truncated by putting the star on an orbit bound to the black hole \citep{hayasaki13}, but this necessarily implies that a larger fraction of the stellar mass is on bound orbits and so should enhance the total energy release. On the other hand, disruptions of giants on parabolic orbits have the same asymptotic $t^{-5/3}$ power law for the rate of return of material to pericenter, but they have a far higher peak.  At its simplest, a constant density spherical star has a return rate proportional to $P^{-5/3}(1-(P/P_0)^{4/3})$ where $P_0$ is the period at the surface, while a shell has a rate simply proportional to $P^{-5/3}$ for $P>P_0$ in both cases.  As a result, disruptions of giant envelopes should show both faster rises and declines.

The fact that ASAS-SN has discovered a likely TDE in the first year of real-time operations, despite only having two cameras operational during most of this period, indicates that we may find these events on a yearly basis now that we are operational in both the Northern and Southern hemispheres. Using the estimates for the local density of black holes from \citet{shankar13} and an observable volume of $3\times10^7$~Mpc$^3$ expected for the fully operational (both hemispheres) ASAS-SN gives a total of $\sim3\times10^4$ black holes with masses in the $M_{BH}=10^6-10^7$~{\msun} mass range that will be observable by ASAS-SN. Given a TDE rate of $(1.5 - 2.0)_{-1.3}^{+2.7} \times 10^{-5}~{\rm yr}^{-1}$ per galaxy \citep{velzen14}, and assuming a 50\% detection efficiency, we calculate an expected rate of roughly $0.1 - 0.7$ TDEs per year to be found in ASAS-SN data. Using the volumetric TDE rate from the same manuscript of $(4 - 8) \times 10^{-8\pm0.4}\,{\rm yr}^{-1}{\rm Mpc}^{-3}$ gives a more optimistic rate of $0.3 - 3$ TDEs per year to be found in ASAS-SN data. These nearby events will be easily observable with a variety of telescopes and instruments, allowing us to study these TDEs across a broad wavelength range and at a level that cannot be done for TDEs discovered at higher redshifts. This will allow us to develop a catalog of well-observed TDE candidates that can be used for population studies, and if candidates are caught early enough, to examine the early behavior of these events. Although ASAS-SN is focused primarily on the discovery and observation of supernovae, it will also be an invaluable resource for studying TDEs and other bright transients in the future.

\section*{Acknowledgments}

The authors thank A. Piro, F. Shankar, N. McConnell, M. Strauss, J. Greene, and J. Guillochon for useful comments and suggestions. We thank M. L. Edwards and the staff of the LBT Observatory for their support and assistance in obtaining the MODS spectra. We thank PI Neil Gehrels and the {\swift} ToO team for promptly approving and executing our observations. We thank LCOGT and its staff for their continued support of ASAS-SN.

Development of ASAS-SN has been supported by NSF grant AST-0908816 and the Center for Cosmology and AstroParticle Physics at the Ohio State University. JFB is supported by NSF grant PHY-1101216.

This research has made use of the XRT Data Analysis Software (XRTDAS) developed under the responsibility of the ASI Science Data Center (ASDC), Italy. At Penn State the NASA {\swift} program is support through contract NAS5-00136.

The LBT is an international collaboration among institutions in the United States, Italy and Germany. LBT Corporation partners are: The University of Arizona on behalf of the Arizona university system; Istituto Nazionale di Astrofisica, Italy; LBT Beteiligungsgesellschaft, Germany, representing the Max-Planck Society, the Astrophysical Institute Potsdam, and Heidelberg University; the Ohio State University, and The Research Corporation, on behalf of The University of Notre Dame, University of Minnesota and University of Virginia.

This publication used data obtained with the MODS spectrographs built with funding from NSF grant AST-9987045 and the NSF Telescope System Instrumentation Program (TSIP), with additional funds from the Ohio Board of Regents and the Ohio State University Office of Research.

The Liverpool Telescope is operated on the island of La Palma by Liverpool John Moores University in the Spanish Observatorio del Roque de los Muchachos of the Instituto de Astrofisica de Canarias with financial support from the UK Science and Technology Facilities Council.

This research was made possible through the use of the AAVSO Photometric All-Sky Survey (APASS), funded by the Robert Martin Ayers Sciences Fund.

Funding for SDSS-III has been provided by the Alfred P. Sloan Foundation, the Participating Institutions, the National Science Foundation, and the U.S. Department of Energy Office of Science. The SDSS-III web site is http://www.sdss3.org/.

SDSS-III is managed by the Astrophysical Research Consortium for the Participating Institutions of the SDSS-III Collaboration including the University of Arizona, the Brazilian Participation Group, Brookhaven National Laboratory, Carnegie Mellon University, University of Florida, the French Participation Group, the German Participation Group, Harvard University, the Instituto de Astrofisica de Canarias, the Michigan State/Notre Dame/JINA Participation Group, Johns Hopkins University, Lawrence Berkeley National Laboratory, Max Planck Institute for Astrophysics, Max Planck Institute for Extraterrestrial Physics, New Mexico State University, New York University, Ohio State University, Pennsylvania State University, University of Portsmouth, Princeton University, the Spanish Participation Group, University of Tokyo, University of Utah, Vanderbilt University, University of Virginia, University of Washington, and Yale University.

This publication makes use of data products from the Two Micron All Sky Survey, which is a joint project of the University of Massachusetts and the Infrared Processing and Analysis Center/California Institute of Technology, funded by the National Aeronautics and Space Administration and the National Science Foundation.

This publication makes use of data products from the Wide-field Infrared Survey Explorer, which is a joint project of the University of California, Los Angeles, and the Jet Propulsion Laboratory/California Institute of Technology, funded by the National Aeronautics and Space Administration.

This research has made use of the NASA/IPAC Extragalactic Database (NED), which is operated by the Jet Propulsion Laboratory, California Institute of Technology, under contract with the National Aeronautics and Space Administration.


\appendix
\section{Follow-up Photometry}
All follow-up photometry is presented in Table~\ref{table:phot} below. Photometry is presented in the natural system for each filter: $ugriz$ magnitudes are in the AB system, while {\swift} filter magnitudes are in the Vega system.

\begin{table*}
\begin{center}
\caption{Photometric data of {\name}. Magnitudes are presented in the natural system for each filter: $ugriz$ magnitudes are in the AB system, while {\swift} filter magnitudes are in the Vega system. \hfill}
\label{table:phot}
\begin{tabular}{lcccc}
\noalign{\medskip}
\hline
MJD & Filter & Magnitude & Magnitude Uncertainty & Telescope \\
\hline
56686.07742 & $z$ & 16.147 & 0.022 & LT \\ 
56694.00965 & $z$ & 16.145 & 0.027 & LT \\ 
56695.05779 & $z$ & 16.149 & 0.025 & LT \\ 
56696.15246 & $z$ & 16.141 & 0.027 & LT \\ 
56697.01584 & $z$ & 16.130 & 0.033 & LT \\ 
56698.03157 & $z$ & 16.120 & 0.026 & LT \\ 
56698.97949 & $z$ & 16.115 & 0.028 & LT \\ 
56699.95359 & $z$ & 16.146 & 0.023 & LT \\ 
56700.97094 & $z$ & 16.135 & 0.030 & LT \\ 
56701.93490 & $z$ & 16.146 & 0.027 & LT \\ 
56710.02390 & $z$ & 16.190 & 0.028 & LT \\ 
56711.04107 & $z$ & 16.210 & 0.023 & LT \\ 
56712.15423 & $z$ & 16.254 & 0.033 & LT \\ 
56713.07523 & $z$ & 16.228 & 0.026 & LT \\ 
56715.02036 & $z$ & 16.264 & 0.027 & LT \\ 
56721.88313 & $z$ & 16.325 & 0.029 & LT \\ 
56723.86738 & $z$ & 16.326 & 0.035 & LT \\ 
56728.01997 & $z$ & 16.337 & 0.030 & LT \\ 
56731.89773 & $z$ & 16.353 & 0.032 & LT \\ 
56733.90845 & $z$ & 16.379 & 0.027 & LT \\ 
56735.92106 & $z$ & 16.335 & 0.034 & LT \\ 
56739.98927 & $z$ & 16.381 & 0.029 & LT \\ 
56741.89896 & $z$ & 16.428 & 0.030 & LT \\ 
56743.98572 & $z$ & 16.402 & 0.030 & LT \\ 
56751.94146 & $z$ & 16.428 & 0.031 & LT \\ 
56753.94364 & $z$ & 16.431 & 0.028 & LT \\ 
56755.04548 & $z$ & 16.419 & 0.031 & LT \\ 
56761.90671 & $z$ & 16.450 & 0.058 & LT \\ 
56762.90040 & $z$ & 16.439 & 0.031 & LT \\ 
56768.92797 & $z$ & 16.438 & 0.035 & LT \\ 
56770.00931 & $z$ & 16.432 & 0.030 & LT \\ 
56770.95330 & $z$ & 16.445 & 0.025 & LT \\ 
56771.97739 & $z$ & 16.481 & 0.028 & LT \\
\hline
56684.12208 & $i$ & 16.245 & 0.022 & LT \\ 
56685.08659 & $i$ & 16.237 & 0.018 & LT \\ 
56686.07649 & $i$ & 16.199 & 0.020 & LT \\ 
56689.38424 & $i$ & 16.211 & 0.036 & LCOGT \\ 
56692.20445 & $i$ & 16.219 & 0.097 & LCOGT \\ 
56694.00872 & $i$ & 16.202 & 0.021 & LT \\ 
56695.05687 & $i$ & 16.195 & 0.022 & LT \\ 
56696.15155 & $i$ & 16.189 & 0.023 & LT \\ 
56696.17710 & $i$ & 16.217 & 0.047 & LCOGT \\ 
56697.01491 & $i$ & 16.192 & 0.026 & LT \\ 
56698.03063 & $i$ & 16.200 & 0.021 & LT \\ 
56698.97856 & $i$ & 16.201 & 0.021 & LT \\ 
56699.95267 & $i$ & 16.196 & 0.024 & LT \\ 
56700.97002 & $i$ & 16.203 & 0.025 & LT \\ 
56701.93398 & $i$ & 16.206 & 0.025 & LT \\ 
56710.02297 & $i$ & 16.283 & 0.021 & LT \\ 
56711.04013 & $i$ & 16.275 & 0.021 & LT \\ 
56712.15330 & $i$ & 16.328 & 0.025 & LT \\ 
56713.07431 & $i$ & 16.327 & 0.021 & LT \\ 
56715.01944 & $i$ & 16.352 & 0.021 & LT \\ 
56721.88220 & $i$ & 16.420 & 0.021 & LT \\ 
56723.86644 & $i$ & 16.427 & 0.023 & LT \\ 
56728.01904 & $i$ & 16.487 & 0.024 & LT \\ 
56731.89681 & $i$ & 16.514 & 0.026 & LT \\ 
56733.90752 & $i$ & 16.493 & 0.025 & LT \\ 
56735.92014 & $i$ & 16.508 & 0.025 & LT \\ 
56739.98834 & $i$ & 16.520 & 0.022 & LT \\ 
56741.89803 & $i$ & 16.590 & 0.022 & LT \\ 
56743.98480 & $i$ & 16.570 & 0.023 & LT \\ 
\hline
\noalign{\smallskip}
\end{tabular}
\end{center}
\end{table*}

\begin{table*}
\setcounter{table}{0}
\begin{center}
\caption{continued. \hfil}
\begin{tabular}{lcccc}
\noalign{\medskip}
\hline
MJD & Filter & Magnitude & Magnitude Uncertainty & Telescope \\
\hline
56751.94053 & $i$ & 16.565 & 0.025 & LT \\ 
56753.94272 & $i$ & 16.577 & 0.024 & LT \\ 
56755.04456 & $i$ & 16.448 & 0.024 & LT \\ 
56761.90579 & $i$ & 16.618 & 0.026 & LT \\ 
56762.89948 & $i$ & 16.597 & 0.024 & LT \\ 
56768.92704 & $i$ & 16.609 & 0.023 & LT \\ 
56770.00838 & $i$ & 16.614 & 0.026 & LT \\ 
56770.95238 & $i$ & 16.608 & 0.021 & LT \\ 
\hline
56684.12117 & $r$ & 16.386 & 0.020 & LT \\ 
56685.08567 & $r$ & 16.355 & 0.021 & LT \\ 
56686.07558 & $r$ & 16.340 & 0.019 & LT \\ 
56689.38237 & $r$ & 16.340 & 0.030 & LCOGT \\ 
56692.20258 & $r$ & 16.374 & 0.045 & LCOGT \\ 
56694.00780 & $r$ & 16.369 & 0.022 & LT \\ 
56695.05595 & $r$ & 16.369 & 0.020 & LT \\ 
56696.15063 & $r$ & 16.371 & 0.020 & LT \\ 
56696.17523 & $r$ & 16.362 & 0.034 & LCOGT \\ 
56697.01400 & $r$ & 16.362 & 0.023 & LT \\ 
56698.02972 & $r$ & 16.345 & 0.021 & LT \\ 
56698.97764 & $r$ & 16.401 & 0.022 & LT \\ 
56699.95175 & $r$ & 16.374 & 0.021 & LT \\ 
56700.96910 & $r$ & 16.412 & 0.027 & LT \\ 
56701.93306 & $r$ & 16.429 & 0.022 & LT \\ 
56710.02204 & $r$ & 16.530 & 0.021 & LT \\ 
56711.03922 & $r$ & 16.539 & 0.020 & LT \\ 
56712.15239 & $r$ & 16.579 & 0.018 & LT \\ 
56713.07339 & $r$ & 16.574 & 0.020 & LT \\ 
56715.01852 & $r$ & 16.635 & 0.020 & LT \\ 
56721.88128 & $r$ & 16.708 & 0.021 & LT \\ 
56723.86553 & $r$ & 16.708 & 0.021 & LT \\ 
56728.01812 & $r$ & 16.762 & 0.024 & LT \\ 
56731.89589 & $r$ & 16.821 & 0.027 & LT \\ 
56733.90660 & $r$ & 16.808 & 0.024 & LT \\ 
56735.91922 & $r$ & 16.830 & 0.022 & LT \\ 
56739.98740 & $r$ & 16.865 & 0.021 & LT \\ 
56741.89711 & $r$ & 16.876 & 0.022 & LT \\ 
56743.98387 & $r$ & 16.839 & 0.022 & LT \\ 
56751.93961 & $r$ & 16.879 & 0.022 & LT \\ 
56753.94180 & $r$ & 16.883 & 0.022 & LT \\ 
56755.04360 & $r$ & 16.907 & 0.023 & LT \\ 
56761.90486 & $r$ & 16.886 & 0.029 & LT \\ 
56762.89856 & $r$ & 16.936 & 0.026 & LT \\ 
56768.92612 & $r$ & 16.930 & 0.020 & LT \\ 
56770.00746 & $r$ & 16.947 & 0.021 & LT \\ 
56770.95146 & $r$ & 16.943 & 0.022 & LT \\ 
56771.97555 & $r$ & 16.968 & 0.020 & LT \\
\hline
56682.51157 & $V$ & 16.300 & 0.100 & ASAS-SN \\ 
56684.90883 & $V$ & 16.420 & 0.041 & {\swift} \\ 
56687.03884 & $V$ & 16.330 & 0.100 & {\swift} \\ 
56697.70773 & $V$ & 16.540 & 0.081 & {\swift} \\ 
56702.77099 & $V$ & 16.390 & 0.110 & {\swift} \\ 
56707.92641 & $V$ & 16.530 & 0.130 & {\swift} \\ 
56729.12832 & $V$ & 17.070 & 0.170 & {\swift} \\ 
56734.38386 & $V$ & 16.730 & 0.170 & {\swift} \\ 
56738.53238 & $V$ & 17.230 & 0.120 & {\swift} \\ 
56739.39868 & $V$ & 17.070 & 0.110 & {\swift} \\ 
56744.18546 & $V$ & 17.180 & 0.170 & {\swift} \\ 
56749.78202 & $V$ & 16.940 & 0.170 & {\swift} \\ 
56755.31624 & $V$ & 17.290 & 0.180 & {\swift} \\ 
56760.19089 & $V$ & 17.160 & 0.150 & {\swift} \\ 
56763.39081 & $V$ & 17.000 & 0.150 & {\swift} \\ 
56770.05206 & $V$ & 17.470 & 0.260 & {\swift} \\ 
\hline
\noalign{\smallskip}
\end{tabular}
\end{center}
\end{table*}

\begin{table*}
\setcounter{table}{0}
\begin{center}
\caption{continued. \hfil}
\begin{tabular}{lcccc}
\noalign{\medskip}
\hline
MJD & Filter & Magnitude & Magnitude Uncertainty & Telescope \\
\hline
56825.86171 & $V$ & 17.620 & 0.240 & {\swift} \\
56828.44541 & $V$ & 17.080 & 0.150 & {\swift} \\
\hline
56684.12022 & $g$ & 16.533 & 0.023 & LT \\ 
56684.33405 & $g$ & 16.528 & 0.036 & LCOGT \\ 
56685.08472 & $g$ & 16.516 & 0.023 & LT \\ 
56686.07463 & $g$ & 16.509 & 0.021 & LT \\ 
56689.38049 & $g$ & 16.503 & 0.027 & LCOGT \\ 
56692.20070 & $g$ & 16.595 & 0.036 & LCOGT \\ 
56694.00685 & $g$ & 16.590 & 0.022 & LT \\ 
56695.05500 & $g$ & 16.594 & 0.022 & LT \\ 
56696.14969 & $g$ & 16.575 & 0.023 & LT \\ 
56696.17336 & $g$ & 16.559 & 0.041 & LT \\ 
56697.01305 & $g$ & 16.597 & 0.025 & LT \\ 
56698.02878 & $g$ & 16.617 & 0.020 & LT \\ 
56698.97669 & $g$ & 16.602 & 0.022 & LT \\ 
56699.95080 & $g$ & 16.633 & 0.029 & LT \\ 
56700.96816 & $g$ & 16.638 & 0.022 & LT \\ 
56701.93212 & $g$ & 16.636 & 0.041 & LT \\ 
56710.02110 & $g$ & 16.837 & 0.022 & LT \\ 
56711.03827 & $g$ & 16.844 & 0.022 & LT \\ 
56712.15143 & $g$ & 16.885 & 0.023 & LT \\ 
56713.07244 & $g$ & 16.897 & 0.021 & LT \\ 
56715.01756 & $g$ & 16.944 & 0.022 & LT \\ 
56721.88033 & $g$ & 17.077 & 0.022 & LT \\ 
56723.86458 & $g$ & 17.104 & 0.024 & LT \\ 
56728.01718 & $g$ & 17.157 & 0.025 & LT \\ 
56731.89495 & $g$ & 17.187 & 0.037 & LT \\ 
56733.90565 & $g$ & 17.233 & 0.029 & LT \\ 
56735.91827 & $g$ & 17.266 & 0.023 & LT \\ 
56739.98646 & $g$ & 17.296 & 0.023 & LT \\ 
56741.89617 & $g$ & 17.353 & 0.024 & LT \\ 
56743.98291 & $g$ & 17.348 & 0.025 & LT \\ 
56751.93865 & $g$ & 17.378 & 0.024 & LT \\ 
56753.94086 & $g$ & 17.379 & 0.024 & LT \\ 
56755.04266 & $g$ & 17.402 & 0.025 & LT \\ 
56761.90391 & $g$ & 17.347 & 0.035 & LT \\ 
56762.89761 & $g$ & 17.436 & 0.033 & LT \\ 
56768.92517 & $g$ & 17.452 & 0.025 & LT \\ 
56770.00652 & $g$ & 17.445 & 0.025 & LT \\ 
56770.95051 & $g$ & 17.442 & 0.022 & LT \\ 
56771.97461 & $g$ & 17.468 & 0.022 & LT \\
56774.95044 & $g$ & 17.466 & 0.021 & LT \\
56778.95189 & $g$ & 17.470 & 0.021 & LT \\ 
56781.91865 & $g$ & 17.481 & 0.022 & LT \\ 
56784.92683 & $g$ & 17.487 & 0.024 & LT \\ 
56789.92206 & $g$ & 17.498 & 0.029 & LT \\
56792.93450 & $g$ & 17.495 & 0.025 & LT \\
56796.90488 & $g$ & 17.522 & 0.022 & LT \\
56799.89040 & $g$ & 17.500 & 0.022 & LT \\
56802.96398 & $g$ & 17.525 & 0.023 & LT \\
\hline
56684.89943 & $B$ & 16.690 & 0.045 & {\swift} \\ 
56687.03569 & $B$ & 16.720 & 0.073 & {\swift} \\ 
56697.70161 & $B$ & 16.830 & 0.054 & {\swift} \\ 
56702.76816 & $B$ & 16.800 & 0.082 & {\swift} \\ 
56707.84216 & $B$ & 17.170 & 0.191 & {\swift} \\ 
56729.12580 & $B$ & 17.480 & 0.112 & {\swift} \\ 
56734.38200 & $B$ & 17.710 & 0.151 & {\swift} \\ 
56738.52596 & $B$ & 17.740 & 0.082 & {\swift} \\ 
56739.39238 & $B$ & 17.740 & 0.082 & {\swift} \\ 
56744.18244 & $B$ & 17.650 & 0.122 & {\swift} \\ 
56749.78476 & $B$ & 17.720 & 0.141 & {\swift} \\ 
56755.31990 & $B$ & 17.510 & 0.112 & {\swift} \\ 
\hline
\noalign{\smallskip}
\end{tabular}
\end{center}
\end{table*}

\begin{table*}
\setcounter{table}{0}
\begin{center}
\caption{continued. \hfil}
\begin{tabular}{lcccc}
\noalign{\medskip}
\hline
MJD & Filter & Magnitude & Magnitude Uncertainty & Telescope \\
\hline
56760.18702 & $B$ & 17.900 & 0.122 & {\swift} \\
56763.38771 & $B$ & 17.570 & 0.112 & {\swift} \\ 
56770.05016 & $B$ & 17.540 & 0.141 & {\swift} \\ 
56825.85845 & $B$ & 17.930 & 0.141 & {\swift} \\
56828.44126 & $B$ & 18.340 & 0.201 & {\swift} \\
\hline
56686.07863 & $u$ & 16.630 & 0.026 & LT \\ 
56694.01086 & $u$ & 16.817 & 0.029 & LT \\ 
56695.05900 & $u$ & 16.848 & 0.015 & LT \\ 
56696.15368 & $u$ & 16.820 & 0.038 & LT \\ 
56697.01705 & $u$ & 16.873 & 0.028 & LT \\ 
56698.03279 & $u$ & 16.896 & 0.024 & LT \\ 
56698.98070 & $u$ & 16.848 & 0.027 & LT \\ 
56699.95480 & $u$ & 16.965 & 0.037 & LT \\ 
56700.97215 & $u$ & 16.928 & 0.070 & LT \\ 
56701.93611 & $u$ & 17.001 & 0.072 & LT \\ 
56710.02511 & $u$ & 17.259 & 0.018 & LT \\ 
56711.04229 & $u$ & 17.240 & 0.040 & LT \\ 
56712.15544 & $u$ & 17.290 & 0.031 & LT \\ 
56713.07644 & $u$ & 17.302 & 0.036 & LT \\ 
56715.02158 & $u$ & 17.411 & 0.033 & LT \\ 
56721.88434 & $u$ & 17.740 & 0.054 & LT \\ 
56723.86859 & $u$ & 17.658 & 0.047 & LT \\ 
56728.02118 & $u$ & 17.817 & 0.056 & LT \\ 
56731.89894 & $u$ & 18.025 & 0.108 & LT \\ 
56733.90965 & $u$ & 17.881 & 0.050 & LT \\ 
56735.92227 & $u$ & 17.992 & 0.046 & LT \\ 
56739.99049 & $u$ & 18.123 & 0.027 & LT \\ 
56741.90017 & $u$ & 18.242 & 0.038 & LT \\ 
56743.98693 & $u$ & 18.260 & 0.042 & LT \\ 
56751.94456 & $u$ & 18.362 & 0.039 & LT \\ 
56753.94485 & $u$ & 18.337 & 0.051 & LT \\ 
56755.04764 & $u$ & 18.372 & 0.048 & LT \\ 
56761.90792 & $u$ & 18.471 & 0.111 & LT \\ 
56762.90161 & $u$ & 18.625 & 0.102 & LT \\ 
56768.92918 & $u$ & 18.604 & 0.060 & LT \\ 
56770.01242 & $u$ & 18.584 & 0.045 & LT \\ 
56770.95452 & $u$ & 18.623 & 0.048 & LT \\ 
56771.97860 & $u$ & 18.564 & 0.042 & LT \\
56774.95245 & $u$ & 18.349 & 0.035 & LT \\
56774.95492 & $u$ & 18.639 & 0.045 & LT \\
56778.95389 & $u$ & 18.684 & 0.038 & LT \\
56778.95636 & $u$ & 18.716 & 0.043 & LT \\
56781.92065 & $u$ & 18.669 & 0.043 & LT \\
56781.92313 & $u$ & 18.696 & 0.044 & LT \\
56784.92883 & $u$ & 18.631 & 0.060 & LT \\
56784.93130 & $u$ & 18.715 & 0.064 & LT \\
56789.92406 & $u$ & 18.603 & 0.096 & LT \\
56792.93650 & $u$ & 18.777 & 0.088 & LT \\
56792.93898 & $u$ & 18.712 & 0.084 & LT \\
56796.90687 & $u$ & 18.807 & 0.049 & LT \\
56796.90934 & $u$ & 18.789 & 0.037 & LT \\
56799.89240 & $u$ & 18.842 & 0.047 & LT \\
56799.89487 & $u$ & 18.779 & 0.051 & LT \\
56802.96599 & $u$ & 18.792 & 0.053 & LT \\
56802.96846 & $u$ & 18.816 & 0.056 & LT \\
\hline
56684.89753 & $U$ & 15.580 & 0.045 & {\swift} \\ 
56687.03513 & $U$ & 15.520 & 0.063 & {\swift} \\ 
56697.70056 & $U$ & 15.770 & 0.054 & {\swift} \\ 
56702.76765 & $U$ & 15.910 & 0.073 & {\swift} \\ 
56707.84122 & $U$ & 16.110 & 0.063 & {\swift} \\ 
56729.12535 & $U$ & 17.000 & 0.122 & {\swift} \\ 
56734.38165 & $U$ & 17.060 & 0.151 & {\swift} \\ 
\hline
\noalign{\smallskip}
\end{tabular}
\end{center}
\end{table*}

\begin{table*}
\setcounter{table}{0}
\begin{center}
\caption{continued. \hfil}
\begin{tabular}{lcccc}
\noalign{\medskip}
\hline
MJD & Filter & Magnitude & Magnitude Uncertainty & Telescope \\
\hline
56738.52485 & $U$ & 17.000 & 0.082 & {\swift} \\ 
56739.39129 & $U$ & 17.090 & 0.082 & {\swift} \\ 
56744.18191 & $U$ & 17.270 & 0.132 & {\swift} \\ 
56749.78160 & $U$ & 17.420 & 0.161 & {\swift} \\ 
56755.31569 & $U$ & 17.390 & 0.141 & {\swift} \\ 
56760.18634 & $U$ & 17.380 & 0.122 & {\swift} \\ 
56763.38716 & $U$ & 17.510 & 0.141 & {\swift} \\ 
56770.04981 & $U$ & 17.890 & 0.251 & {\swift} \\ 
56825.85787 & $U$ & 18.120 & 0.211 & {\swift} \\
56828.44054 & $U$ & 18.230 & 0.251 & {\swift} \\
\hline
56684.89377 & $UVW1$ & 15.110 & 0.042 & {\swift} \\ 
56687.03357 & $UVW1$ & 15.090 & 0.050 & {\swift} \\ 
56697.69751 & $UVW1$ & 15.550 & 0.050 & {\swift} \\ 
56702.76625 & $UVW1$ & 15.710 & 0.058 & {\swift} \\ 
56707.83852 & $UVW1$ & 15.910 & 0.050 & {\swift} \\ 
56729.07071 & $UVW1$ & 16.870 & 0.076 & {\swift} \\ 
56734.38074 & $UVW1$ & 16.920 & 0.114 & {\swift} \\ 
56738.52164 & $UVW1$ & 17.090 & 0.076 & {\swift} \\ 
56739.38815 & $UVW1$ & 17.120 & 0.076 & {\swift} \\ 
56744.18042 & $UVW1$ & 17.220 & 0.104 & {\swift} \\ 
56749.78435 & $UVW1$ & 17.360 & 0.133 & {\swift} \\ 
56755.31935 & $UVW1$ & 17.400 & 0.114 & {\swift} \\ 
56760.18441 & $UVW1$ & 17.830 & 0.124 & {\swift} \\ 
56763.38562 & $UVW1$ & 17.820 & 0.133 & {\swift} \\ 
56768.73185 & $UVW1$ & 18.180 & 0.212 & {\swift} \\ 
56770.04886 & $UVW1$ & 18.000 & 0.202 & {\swift} \\
56794.79258 & $UVW1$ & 18.550 & 0.173 & {\swift} \\
56799.45558 & $UVW1$ & 18.150 & 0.163 & {\swift} \\
56804.11631 & $UVW1$ & 18.390 & 0.341 & {\swift} \\
56809.31304 & $UVW1$ & 18.500 & 0.262 & {\swift} \\
56814.44711 & $UVW1$ & 18.220 & 0.232 & {\swift} \\
56825.85625 & $UVW1$ & 18.600 & 0.202 & {\swift} \\
56828.43848 & $UVW1$ & 18.530 & 0.192 & {\swift} \\
\hline
56684.91259 & $UVM2$ & 14.770 & 0.042 & {\swift} \\ 
56687.03940 & $UVM2$ & 14.780 & 0.042 & {\swift} \\ 
56697.70877 & $UVM2$ & 15.430 & 0.042 & {\swift} \\ 
56702.77148 & $UVM2$ & 15.640 & 0.050 & {\swift} \\ 
56707.92682 & $UVM2$ & 15.850 & 0.067 & {\swift} \\ 
56729.12878 & $UVM2$ & 16.880 & 0.076 & {\swift} \\ 
56734.38421 & $UVM2$ & 16.910 & 0.104 & {\swift} \\ 
56738.53348 & $UVM2$ & 17.140 & 0.067 & {\swift} \\ 
56739.39976 & $UVM2$ & 17.000 & 0.058 & {\swift} \\ 
56744.18599 & $UVM2$ & 17.130 & 0.085 & {\swift} \\ 
56749.78046 & $UVM2$ & 17.330 & 0.104 & {\swift} \\ 
56755.31415 & $UVM2$ & 17.600 & 0.104 & {\swift} \\ 
56760.19156 & $UVM2$ & 17.740 & 0.095 & {\swift} \\ 
56763.39135 & $UVM2$ & 17.690 & 0.104 & {\swift} \\ 
56770.05241 & $UVM2$ & 17.710 & 0.133 & {\swift} \\ 
56794.59550 & $UVM2$ & $>18.13$ &  N/A & {\swift} \\ 
56799.37997 & $UVM2$ & 18.250 & 0.182 & {\swift} \\ 
56804.11453 & $UVM2$ & 18.430 & 0.222 & {\swift} \\
56809.30874 & $UVM2$ & 18.600 & 0.153 & {\swift} \\ 
56814.44225 & $UVM2$ & 18.430 & 0.133 & {\swift} \\ 
56825.86227 & $UVM2$ & 18.880 & 0.182 & {\swift} \\ 
56828.44612 & $UVM2$ & 18.570 & 0.133 & {\swift} \\ 
\hline
56684.90137 & $UVW2$ & 15.060 & 0.042 & {\swift} \\ 
56687.03627 & $UVW2$ & 15.170 & 0.050 & {\swift} \\ 
56697.70267 & $UVW2$ & 15.800 & 0.042 & {\swift} \\ 
56702.76867 & $UVW2$ & 15.980 & 0.058 & {\swift} \\ 
56707.92452 & $UVW2$ & 16.030 & 0.058 & {\swift} \\ 
56729.12627 & $UVW2$ & 16.910 & 0.085 & {\swift} \\ 
56734.38236 & $UVW2$ & 17.120 & 0.104 & {\swift} \\ 
\hline
\noalign{\smallskip}
\end{tabular}
\end{center}
\end{table*}

\begin{table*}
\setcounter{table}{0}
\begin{center}
\caption{continued. \hfil}
\begin{tabular}{lcccc}
\noalign{\medskip}
\hline
MJD & Filter & Magnitude & Magnitude Uncertainty & Telescope \\
\hline
56734.38236 & $UVW2$ & 17.120 & 0.104 & {\swift} \\ 
56738.52707 & $UVW2$ & 17.130 & 0.058 & {\swift} \\ 
56739.39347 & $UVW2$ & 17.170 & 0.067 & {\swift} \\ 
56744.18299 & $UVW2$ & 17.230 & 0.085 & {\swift} \\ 
56749.78247 & $UVW2$ & 17.270 & 0.104 & {\swift} \\ 
56755.31680 & $UVW2$ & 17.460 & 0.095 & {\swift} \\ 
56760.18770 & $UVW2$ & 17.690 & 0.095 & {\swift} \\ 
56763.38825 & $UVW2$ & 17.690 & 0.104 & {\swift} \\ 
56770.05052 & $UVW2$ & 17.580 & 0.124 & {\swift} \\ 
56794.59036 & $UVW2$ & 18.470 & 0.104 & {\swift} \\
56799.37892 & $UVW2$ & 18.150 & 0.202 & {\swift} \\
56804.11361 & $UVW2$ & 18.490 & 0.252 & {\swift} \\
56809.30656 & $UVW2$ & 18.230 & 0.143 & {\swift} \\
56814.43980 & $UVW2$ & 18.770 & 0.192 & {\swift} \\
56825.85903 & $UVW2$ & 18.930 & 0.182 & {\swift} \\
56828.44199 & $UVW2$ & 18.650 & 0.143 & {\swift} \\
\hline
\noalign{\smallskip}
\end{tabular}
\end{center}
\end{table*}
\end{document}